\RequirePackage{fix-cm}
\documentclass[smallextended]{svjour4}       
\smartqed  
\usepackage{graphicx}
\usepackage{hyperref}

\usepackage[usenames]{color}

\def\be{\begin{equation}}
\def\ee{\end{equation}}
\def\ba{\begin{eqnarray}}
\def\ea{\end{eqnarray}}

\def\sfrac#1#2{{\textstyle \frac{#1}{#2}}}

\begin{document}

\title{$N^\ast$ Form Factors based on a Covariant Quark Model\footnote[4]{
This is a post-peer-review, pre-copyedit version of an article 
published in \\ {\color{blue}Few-Body Syst, {\bf 59} (2018)}. 
The final authenticated version is available online at: 
{\color{blue}http://doi.org/10.1007/s00601-018-1412-9}.}
}


\author{G.~Ramalho}


\institute{G.~Ramalho \at
              Laborat\'orio de F\'{i}sica Te\'orica e Computacional -- LFTC, \\
               Universidade Cruzeiro do Sul, 01506-000, S\~ao Paulo, SP, Brazil \\
              Tel.: +55 11 3385-3004\\
              \email{gilberto.ramalho@cruzeirodosul.edu.br} 
}

\date{Published date: June 4, 2018}

\maketitle

\begin{abstract}
We discuss the results of the covariant spectator quark model 
for the several $\gamma^\ast N \to N^\ast$ transitions, 
where $N$ is the nucleon and $N^\ast$ a nucleon excitation.
More specifically, we present predictions 
for the form factors and transition amplitudes 
associated with the resonances $N(1440)1/2^+$,
$N(1535)1/2^-$,  $N(1520)3/2^-$, $\Delta(1620)1/2^-$ and $N(1650)1/2^-$.  
The estimates based on valence quark degrees of freedom  
are compared with the available data,
particularly with the recent Jefferson Lab data 
at low and large momentum transfer ($Q^2$).
In general the estimates are in good agreement 
with the empirical data for $Q^2 >2$ GeV$^2$, with a few exceptions.
The results are discussed in terms of 
the role of the valence quarks and 
the meson cloud excitations for the different  resonances $N^\ast$.
We also review our results for the resonance $\Delta(1232)3/2^+$
and discuss the relevance of the pion cloud component 
for the magnetic dipole form factor and 
the electric and Coulomb quadrupole form factors.
\keywords{Nucleon excitations \and Electromagnetic structure \and  
Valence quarks \and Meson cloud}
\end{abstract}

\newpage

\section{Introduction}
\label{secIntro}

Modern accelerators, such as MAMI, MIT-Bates and Jefferson Lab (JLab) 
provide nowadays important information about the 
electromagnetic structure of the nucleon ($N$) 
and nucleon excitation ($N^\ast$)
up to masses of 2 GeV~\cite{NSTAR,Burkert04,Aznauryan12,MAID1,Eichmann16}.
The data associated to the $\gamma^\ast N \to N^\ast$ transition,
for the photon momentum $q$ 
can be represented in terms of structure functions,
transition form factors or helicity  amplitudes, 
dependent on the transition momentum square $q^2$,
or from $Q^2=-q^2$.
In the recent years data associated with several resonances $N^\ast$  
have been collected at JLab up to $Q^2= 6$ GeV$^2$.
With the JLab-12 GeV upgrade, we expect to 
achieve in a near future $Q^2 \simeq 12$ GeV$^2$~\cite{NSTAR}.

To interpret the recent data at the range $Q^2=2$--6 GeV$^2$ 
and above, it is necessary to develop theoretical models based on relativity.
Preferable are models based on the dominant degrees of freedom 
at large $Q^2$, the valence quarks. 
However, models that take into account the degrees of freedom 
associated with the quark-antiquark/meson states
may be also appropriated to understand the 
transition between the low and large $Q^2$ regimes.
These models can be used to make predictions
for transition form factors at large $Q^2$, 
and may also be used to guide future experiments
as the ones projected to the JLab-12 GeV upgrade~\cite{NSTAR,Aznauryan12,Nucleon,NDelta,NDeltaD}.

Different frameworks have been used in the study of the   
$\gamma^\ast N \to N^\ast$ transitions,
such as, quark models, effective chiral perturbation theories, 
dynamical coupled-channel models, Dyson-Schwinger equations,
large $N_c$ limit, QCD sum rules,  
perturbative QCD and lattice QCD simulations, among others~\cite{NSTAR,Burkert04}.
As mentioned, frameworks based on valence quarks  
are particularly useful at large $Q^2$,  since those degrees 
of freedom are expected to be dominant.

In the present work we discuss mainly results based on 
the covariant spectator quark model. 
The  covariant spectator quark model is a model 
based on constituent quarks where the 
quark electromagnetic is parametrized in order 
to describe the nucleon electromagnetic structure~\cite{Nucleon}.
The wave functions of the baryons are ruled by 
the $SU(6)\otimes O(3)$ symmetry, with radial wave functions
determined phenomenologically with the assistance 
of empirical data, lattice data
or estimates of the quark core contributions~\cite{Nucleon,NDelta,NDeltaD,Omega,SQTM,LatticeD}.  
One can then use parametrizations of a few resonances $N^\ast$
to make predictions for other states based on the symmetries.
In this work we present a few examples. 
The model is covariant by construction and therefore can be  used 
at very large $Q^2$.
In some cases the model can be  extended with the inclusion 
of effective descriptions of the meson cloud effects,
that can be significant at small $Q^2$~\cite{NSTAR,Burkert04,Aznauryan12,NDelta,NDeltaD}.
In Sect.~\ref{secDelta} this methodology 
is illustrated for the case of the $\gamma^\ast N \to \Delta(1232)$ transition.
The details about the covariant spectator quark model 
are discussed in the next section (Sect.~\ref{secCSQM}).

In the following sections, we discuss the results 
for several $\gamma^\ast N \to N^\ast$ form factors.
We start with our recent results 
for the $N(1535)1/2^-$ and $N(1520)3/2^-$ resonances 
in the context of the semirelativistic approximation (Sect.~\ref{secSRapp}).
Next, we discuss briefly
the results for the  $N(1440)1/2^+$ based on the covariant spectator quark model 
and also an estimate of the valence quark contributions 
based on holographic methods (Sect.~\ref{secRoper}).
In Sect.~\ref{secDelta}, we review the covariant spectator quark model results 
for the $\Delta(1232)3/2^+$ and discuss also some recent 
results for the quadrupole form factors 
in the light of Siegert's theorem.
We finish our presentation of results
with the estimates of the negative parity states
from the $[70,1^-]$ supermultiplet, 
combining the covariant spectator quark model with the single quark transition model, 
for the cases $N(1650)1/2^-$ and $\Delta(1620)1/2^-$ (Sect.~\ref{secSQTM}).
At the end, we summarize the results and conclusions 
associated with the resonances  $N^\ast$
discussed in the present work (Sect.~\ref{secConclusions}).

\section{Covariant Spectator Quark Model}
\label{secCSQM}

The covariant spectator quark model is 
based on the covariant spectator theory~\cite{Nucleon,Gross,Stadler97}.
The model treats the baryons, 
including the nucleon and the nucleon excitations, 
as three-quark systems~\cite{Nucleon,Nucleon2,NucleonDIS}.
The baryon wave functions are expressed in terms of 
the quarks states, according with the $SU(6)\otimes O(3)$ symmetry group~\cite{Capstick00,Giannini15}.
In the spectator theory the baryon can be regarded 
as an off-mass-shell quark free to couple with the photon fields 
and two spectator on-mass-shell quarks~\cite{Nucleon,Omega,Nucleon2,Octet,OctetMedium}. 
One can integrate over the quark-pair degrees of freedom
and reduce the baryon a quark-diquark system,
where the diquark represents an on-shell spectator 
particle with an effective mass $m_D$~\cite{Nucleon,Omega,Nucleon2}.
At the end, we obtain an effective quark-diquark wave function,
free of singularities which include the quark 
confinement implicitly~\cite{Nucleon,Omega,Stadler97,Nucleon2,Gross06,Savkli01}.

In the covariant spectator quark model 
the electromagnetic interaction is described by the photon coupling 
with the constituent quarks in relativistic impulse approximation.
The structure of the constituent quarks is 
represented by the quark structure form factors 
which encode effectively the gluon and 
quark-antiquark substructure of those quarks~\cite{Nucleon,Omega}.
In the $SU(2)$ flavor sector, one uses the form~\cite{Nucleon}
\ba
j_q^\mu =
\left(\frac{1}{6}f_{1+}  + \frac{1}{2} f_{1-} \tau_3\right) \gamma^\mu
+ \left(\frac{1}{6}f_{2+}  + \frac{1}{2} f_{2-} \tau_3\right)
\frac{i \sigma^{\mu \nu} q_\nu}{2M_N}, 
\label{eqjq}
\ea
where $M_N$ is the nucleon mass 
and $f_{i\pm}$ ($i=1,2$) are the isoscalar/isovector components 
of the Dirac ($i=1$) and Pauli ($i=2$) quark form form factors.
In Eq.~(\ref{eqjq}), $\tau_3$ is the Pauli operator and 
acts on the isospin states 
of the baryons. For more details, check Refs.~\cite{Nucleon,Omega,Octet,OctetMedium}.
The quark current (\ref{eqjq}) can be generalized 
to the $SU(3)$ sector~\cite{Omega,Octet,Omega2} and to the axial-vector case~\cite{Axial}.

For convenience, we label the nucleon resonance $N^\ast$ by $R$.
When the nucleon wave function ($\Psi_N$) 
and the resonance wave function ($\Psi_R$)
are both expressed in terms of
the single quark and quark-pair states,
the transition current in impulse approximation 
can be written as~\cite{Nucleon,Omega,Nucleon2}
\ba
J^\mu=
3 \sum_{\Gamma} 
\int_k \bar \Psi_R (P_R,k) j_q^\mu \Psi_N(P_N,k),
\label{eqJmu}
\ea  
where $P_R$, $P_N$, and $k$ are  
the resonance, the nucleon, and the diquark momenta, respectively.
In the previous equation 
the index $\Gamma$ labels
the intermediate diquark polarization states,
the factor 3 takes account of the contributions from
the other quark pairs by the symmetry, and the integration
symbol represents the covariant integration over the 
diquark on-mass-shell momentum.
In the study of the inelastic transitions 
we use the Landau prescription to ensure
the current conservation~\cite{SQTM,SRapp,N1535,N1520}.

Using Eq.~(\ref{eqJmu}), we can express 
the transition current in terms of the 
quark electromagnetic form factors $f_{i\pm}$ ($i=1,2$)
and the radial wave functions 
$\psi_N$ and $\psi_R$~\cite{Nucleon,SQTM,N1535,N1520}.
The radial wave functions $\psi_B(P,k)$, with $B=N,R$,
are scalar functions that 
depend on the  baryon ($P$) and diquark ($k$) momenta,
and parametrize the momentum distributions 
of the quark-diquark systems.
Since the baryon and the diquark are both on-mass-shell 
the dependence on the momenta can be expressed 
in terms of the dimensionless variable 
$\chi = \frac{(M_B-m_D)^2 - (P-  k)^2}{M_B m_D}$,
where $M_B$ and $m_D$ are the baryon and 
the diquark masses, respectively~\cite{Nucleon,Omega}.

The quark electromagnetic form factors $f_{i\pm}$   
are parametrized according to a vector meson dominance
mechanism~\cite{Nucleon,Omega,LatticeD,Lattice}.
Taking advantage of the quark form factor structure based on vector 
meson dominance,
the model has been extended to the lattice QCD regime
(heavy pions and no meson 
cloud)~\cite{Omega,LatticeD,Octet,Omega2,Lattice},
to the nuclear medium~\cite{OctetMedium} 
and to the timelike regime ($Q^2 < 0$)~\cite{N1520TL,NDeltaTL,NDeltaTL2}.
In the generalization to those regimes we use the representation 
of the radial wave functions $\psi_R$ in terms of the diquark mass ($m_D$) and 
baryon mass ($M_B$)~\cite{LatticeD,Lattice}.

The covariant spectator quark model
was originally developed for the nucleon electromagnetic structure~\cite{Nucleon}.
In this first study it was shown that the electromagnetic 
structure of the nucleon can be described 
based on a calibration of quark electromagnetic form 
factors $f_{i\pm}$ defined by Eq.~(\ref{eqjq}) 
and an appropriated form for the radial wave function $\psi_N(P,k)$.
The parametrization from Ref.~\cite{Nucleon} for the quark current 
and radial wave function have been used in all subsequent calculations 
of the $\gamma^\ast N \to N^\ast$ transition form factors.

The previous description of the covariant spectator quark model 
takes into account only the effects associated 
with the valence quark degrees of freedom.
There are however some processes, such as
the meson exchanged between the different quarks
inside the baryon, which cannot be reduced
to processes associated with the dressing of a single quark.
Those processes can be regarded  as a consequence 
of the meson exchanged between the different quarks inside
the baryon, and can be classified as meson cloud corrections 
to the hadronic reactions~\cite{Octet,N1520,N1520TL,OctetDecuplet,OctetDecuplet2}.
As mentioned the meson cloud effects can be very significant 
at small $Q^2$.

The study of the role of the meson cloud effects 
on the $\gamma^\ast N \to N^\ast$ transition can be done 
also in the context of the 
dynamical coupled-channel reaction models~\cite{Burkert04,EBACreview,JDiaz07}.
Those models use baryon-meson states to 
describe the photo- and electro-production of mesons by nucleons,
taking into account the meson dressing of propagators and vertices. 
Once determined the meson couplings by fits to the data,
the framework can be used to extract indirectly 
the effect of the bare core contribution to the data, 
removing the effect of the meson-baryon dressing of propagators and vertices. 
Those estimates of the bare core can be very useful to test 
the limits of models based on valence quarks, as discussed 
in Sect.~\ref{secDelta}, for the case of the $\Delta(1232)$.
Examples of dynamical coupled-channel reaction models
are the Sato-Lee~\cite{SatoLee}, the DMT~\cite{Kamalov} and the 
EBAC/Argonne-Osaka models~\cite{EBACreview,JDiaz07,EBAC}.

The model generalized to the $SU(3)$-flavor sector 
have been used in the calculation of octet 
and decuplet form factors~\cite{Omega,Octet,Omega2,OctetDecuplet,OctetDecuplet2,Deformation,DeltaFF,DeltaDFF,OctetMM},
and other transition form  factors 
in the spacelike region ($Q^2> 0$)~\cite{SQTM,LambdaStar,Delta1600}.
The model also has been used in studies
of the electromagnetic structure of resonances $N^\ast$ 
in the timelike region ($Q^2 < 0$)~\cite{N1520TL,NDeltaTL,NDeltaTL2}.
Applications of the model to the axial and deep inelastic structure 
of the nucleon can be found in Refs.~\cite{Nucleon,Nucleon2,NucleonDIS,Axial}.

In the following sections 
we discuss the results from the covariant spectator quark model 
for the electromagnetic structure of 
resonances $\Delta(1232)3/2^+$, $N(1440)1/2^+$, $N(1535)1/2^-$,
$N(1520)3/2^-$, $\Delta(1620)1/2^-$ and $N(1650)1/2^-$.

\section{$N(1535)1/2^-$ and $N(1520)3/2^-$}
\label{secSRapp}

We now discuss the negative parity states $N(1535)$ and $N(1520)$.
In this study, we use previous calculations
based on the covariant spectator quark model~\cite{N1535,N1520,N1520TL},
combined with the semirelativistic approximation, described below.

\subsection{Semirelativistic approximation}

As mentioned, the information about the $\gamma^\ast N \to N^\ast$ transitions
can be characterized by transition form factors 
dependent on the invariant $Q^2$.
For the discussion of the kinematics, it is however convenient 
to choose a specific frame.
In the following discussion, we use $|{\bf q}|$ 
to represent the photon three-momentum at the resonance 
$R$ rest frame.

The description of the $\gamma^\ast N \to N^\ast$ transition 
is simplified in a nonrelativistic framework, 
because in that case we do not need to take into account the energy component.
In these conditions, the orthogonality between states
is defined when $|{\bf q}|=0$, and both particles 
are at rest ($E_R \approx E_N \approx M_N$).
As a consequence, the transition form factors are independent 
of the resonance mass~\cite{SRapp}.

In a relativistic framework the discussion become more 
intricate because there are ambiguities 
related with the relativistic generalization of the states. 
One of the problems is the definition the orthogonality 
between states, since the two states cannot be at 
rest in the same frame, unless $M_R=M_N$.
At $Q^2=0$ one can write $|{\bf q}|=\sfrac{M_R^2-M_N^2}{2M_R}$,
therefore, one has  $|{\bf q}|=0$, only in the limit $M_R=M_N$.
These ambiguities difficult the calculation 
of transition amplitudes and helicity amplitudes  in some cases
when $M_R \ne M_N$~\cite{N1535,N1520}.

In the semirelativistic approximation, 
we start by considering the approximation $M_R=M_N$ 
in the calculation of the elementary form factors,
defined precisely in the following sections.
In the next step we use those results
to calculate the multipole form factors
and helicity amplitudes, 
which are defined for $M_R \ne M_N$,
and compare the results with the measured data~\cite{SRapp}.
An important aspect about the semirelativistic approximation 
is the way the radial wave functions $\psi_R$ are defined.
We use a form for $\psi_R$ that allows us to obtain 
parameter free results for the transition form factors 
and the helicity amplitudes,
based on a relation between $\psi_R$ and $\psi_N$.

With the semirelativistic approximation one tries then to achieve two goals.
On the one hand, we want to keep the nice analytic proprieties 
of the form factors which are spoiled in the relativistic generalization 
of the wave functions in the case $M_R \ne M_N$.
On the other hand, we want to describe the experimental 
helicity amplitudes, which are defined only in the case  $M_R \ne M_N$~\cite{SRapp}.

\subsection*{Notation}  

Before discussing the $N(1535)$ and $N(1520)$ cases 
it is convenient to introduce some general notation. 

To represent the transition form factors we 
use the symmetric ($S$) and anti-symmetric ($A$) 
combination of quark currents, 
which are expressed as a 
combination of quark form factors~\cite{Nucleon,Octet,OctetMedium} ($i=1,2$):
\ba
j_i^S= \frac{1}{6} f_{i+} +  \frac{1}{2} f_{i-} \tau_3,
\hspace{1.5cm}
j_i^A= \frac{1}{6} f_{i+} -  \frac{1}{6} f_{i-} \tau_3.
\ea

It is also convenient to 
consider the following overlap integral
\ba
{\cal I}_R (Q^2)= \int_k \frac{k_z}{|{\bf k}|} \psi_R (P_R,k) \psi_N(P_N,k),
\label{eqIR}
\ea
where $P_R$ and $P_N$ are the momentum of the 
resonance and nucleon respectively.
For simplicity, we expressed 
the integral (\ref{eqIR}) at the resonance rest frame,
but it can be generalized to an arbitrary frame~\cite{SRapp,N1535,N1520}.

An important characteristic of the semirelativistic approximation is that 
we assume that the radial wave functions of the resonances 
can be defined in the rest frame 
by the nucleon wave function, $\psi_R \equiv \psi_N$.
As a consequence of this assumption, one has~\cite{SRapp} 
\ba
{\cal I}_R \propto |{\bf q}|,
\label{eqIR1}
\ea 
in the limit $M_R \to M_N$.

The previous equation implies that 
the resonance $R$ and the nucleon are orthogonal states, 
since in the limit $Q^2=0$, one has $|{\bf q}| =0$.
Another consequence of Eq.~(\ref{eqIR1}) is that
the final results are independent of the 
radial structure of the resonance $R$ and 
depend only on the parametrization  of $\psi_N$.
As a consequence, the estimates based on the 
semirelativistic approximation have no free 
parameters, apart the parameters used on $\psi_N$,
and then provide true predictions for the transition form factors and 
helicity amplitudes.

In the following, we implement the semirelativistic limit 
replacing the dependence on $M_R$ and $M_N$ by the $M$,
where $M \equiv \sfrac{1}{2}(M_N + M_R)$.
Using this notation, we can 
write $|{\bf q}| = Q \sqrt{1 + \tau}$, with $\tau= \frac{Q^2}{4 M^2}$.

\subsection{$N(1535)1/2^-$}

The $\gamma^\ast N \to N(1535)$ transition current 
can be expressed using units of elementary charge ($e$), in the form  
\ba
J^\mu = \bar u_R
\left[F_1^\ast 
\left( \gamma^\mu -  \frac{{\not \! q} q^\mu}{q^2}
\right)
 + F_2^\ast \frac{i \sigma^{\mu \nu} q_\mu}{M_R + M_N} \right] \gamma_5 u_N,
\label{eqJS11}
\ea
where $u_R$ and $u_N$ are the resonance and nucleon spinors, respectively.
Equation (\ref{eqJS11}) defines the elementary form factors,
Dirac ($F_1^\ast$) and Pauli ($F_2^\ast$)~\cite{NSTAR,SRapp,N1535}.

\begin{figure}[t]
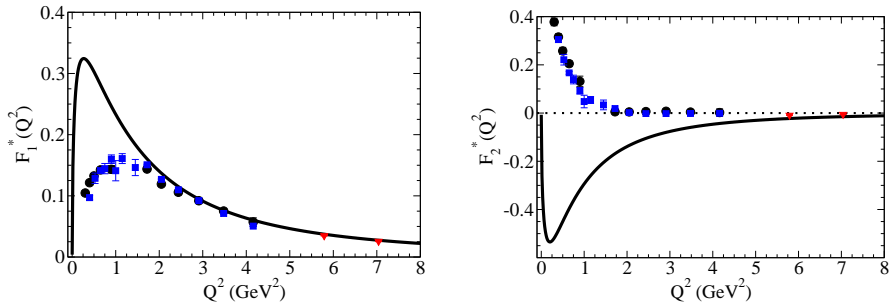
 
\vspace{.5cm}
\includegraphics[width=5.5cm]{F1_mod3B1}  \hspace{.4cm}
\includegraphics[width=5.5cm]{F2_mod3B1} 
\caption{$\gamma^\ast N \to N(1535)$ transition form factors.
Data from  MAID~\cite{MAID1}, 
CLAS~\cite{Aznauryan09} and  JLab/Hall C~\cite{Dalton09}. 
}
\label{figN1535}       
\end{figure}

In the semirelativistic limit, we obtain the following results~\cite{SRapp}:
\ba
F_1^\ast= \frac{1}{2}(3 j_1^S + j_1^A) {\cal I}_R,
\hspace{1.cm}
F_2^\ast= - \frac{1}{2}(3 j_2^S - j_2^A) {\cal I}_R.
\ea 
The numerical results for the form factors are presented in Fig.~\ref{figN1535},
in comparison with the data from Refs.~\cite{MAID1,Aznauryan09,Dalton09}.

Both form factors vanish at $Q^2=0$ as a consequence of the relation 
$F_i^\ast (Q^2)  \propto |{\bf q}|$. 
For the Dirac form factor we obtain a good description 
of the data for $Q^2 > 2$ GeV$^2$.
As for $F_2^\ast$ the model fails to describe the sign of the data.
We can notice in addition that in the $Q^2 > 2$ GeV$^2$ region 
the experimental value of the Pauli form factor 
is compatible with zero ($F_2^\ast \simeq 0$).

The failure of the semirelativistic approximation 
for the Dirac form factor below 2 GeV$^2$, and for the Pauli form factor
can be interpreted as a consequence of the omission of 
the meson cloud effects~\cite{SRapp}. 
Our results for both form factors compare well 
with a previous estimate of the bare core contributions 
based on the EBAC coupled-channel dynamical model~\cite{EBAC}.
Our present results can be tested in a near future by 
new estimates based on the Argonne-Osaka model~\cite{Kamano-talk}.

The discussion about the implications of the result $F_2^\ast \simeq 0$, and 
the possible physical interpretations 
are presented at the end of the section.

As a consequence of the model results for $F_2^\ast$,
the model estimates are not comparable with the experimental helicity amplitudes,
except in some specific limits.
A more detailed discussion can be found in Ref.~\cite{SRapp}.

\subsection{$N(1520)1/2^-$}

The $\gamma^\ast N \to N(1520)$ transition current 
can be expressed, in units $e$, as~\cite{N1520,N1520TL}:
\ba
J^\mu = 
\bar u_\alpha \left[
G_1 \, q^\alpha \gamma^\mu + G_2 \, q^\alpha P^\mu + G_3 \, q^\alpha q^\mu
+ ... \right] u_N
\ea
where $u_\alpha$ is the Rarita-Schwinger of the $R$ state,
$u_N$ is the nucleon spinor, $P=\sfrac{1}{2}(P_R + P_N)$, 
and the dots indicate 
gauge terms that are not relevant to the present discussion.
The functions $G_i$ ($i=1,2,3$) are the elementary form factors of the transition. 

The results for the elementary form factors in the 
semirelativistic approximation are~\cite{SRapp}:
\ba
& &
G_1= - \frac{3}{2 \sqrt{2}} 
\left[ \left(j_1^A + \frac{1}{3} j_1^S \right)
+ \left(j_2^A + \frac{1}{3} j_2^S \right)
\right] \frac{{\cal I}_R}{|{\bf q}|} 
\label{eqG1}\\
& &
G_2= + \frac{3}{2 \sqrt{2} M} 
\left[ j_2^A + \frac{1}{3}\frac{1- 3 \tau}{1 + \tau} j_2^S
+ \frac{4}{3} j_1^S
\right] \frac{{\cal I}_R}{|{\bf q}|} 
\label{eqG2}\\
& &
G_3 =0,
\label{eqG3}
\ea
where $\tau= \frac{Q^2}{(M_R + M_N)^2} \equiv \frac{Q^2}{4 M^2}$.

For the purpose of the discussion, we note that 
$G_1$ and $G_2$ are proportional to $\frac{{\cal I}_R}{|{\bf q}|}$
and are therefore well defined at the photon point,
according with Eq.~(\ref{eqIR1}).

\begin{figure}[t]
\vspace{.5cm}
\mbox{
\includegraphics[width=1.45in]{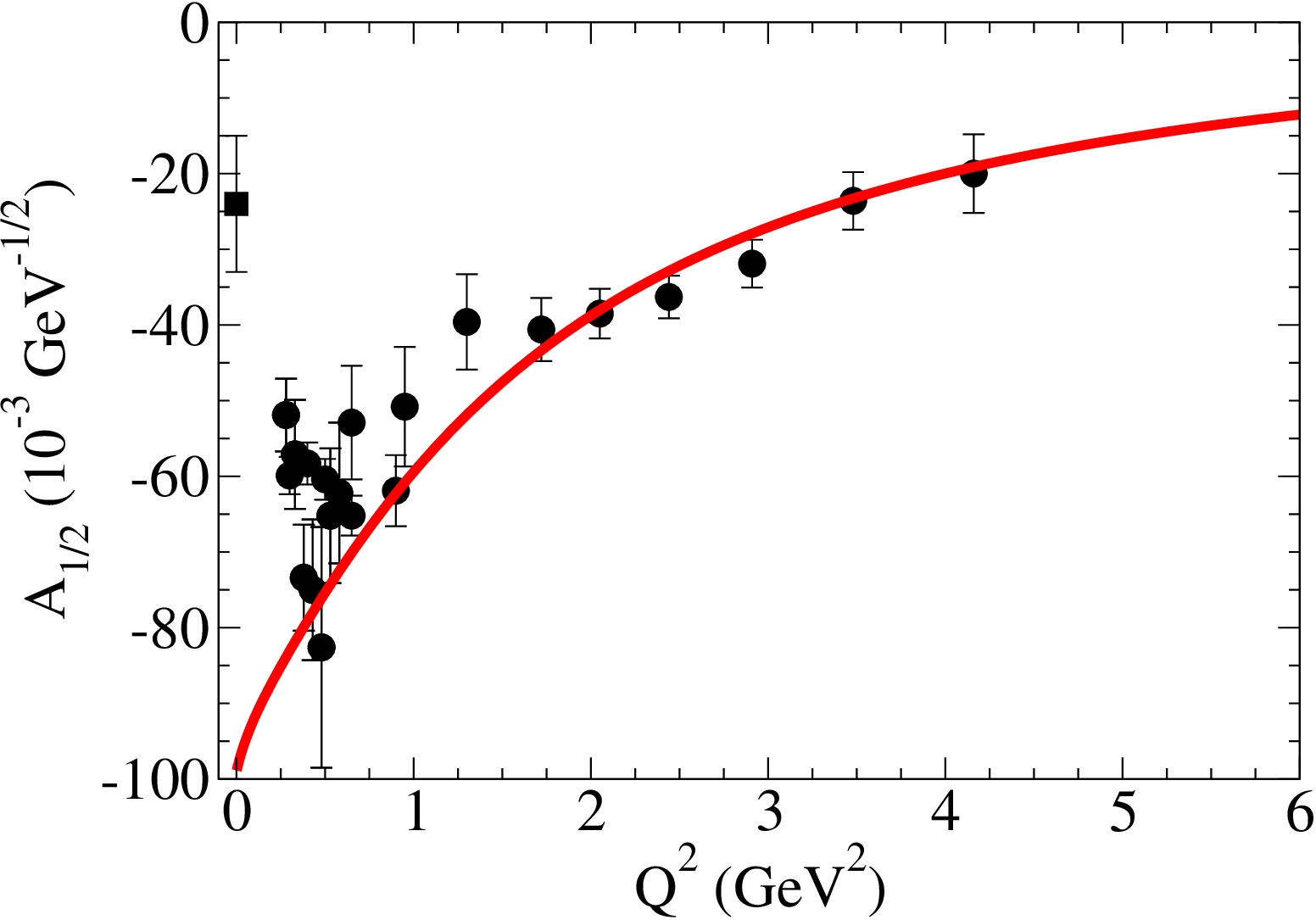}  \hspace{.1cm}
\includegraphics[width=1.45in]{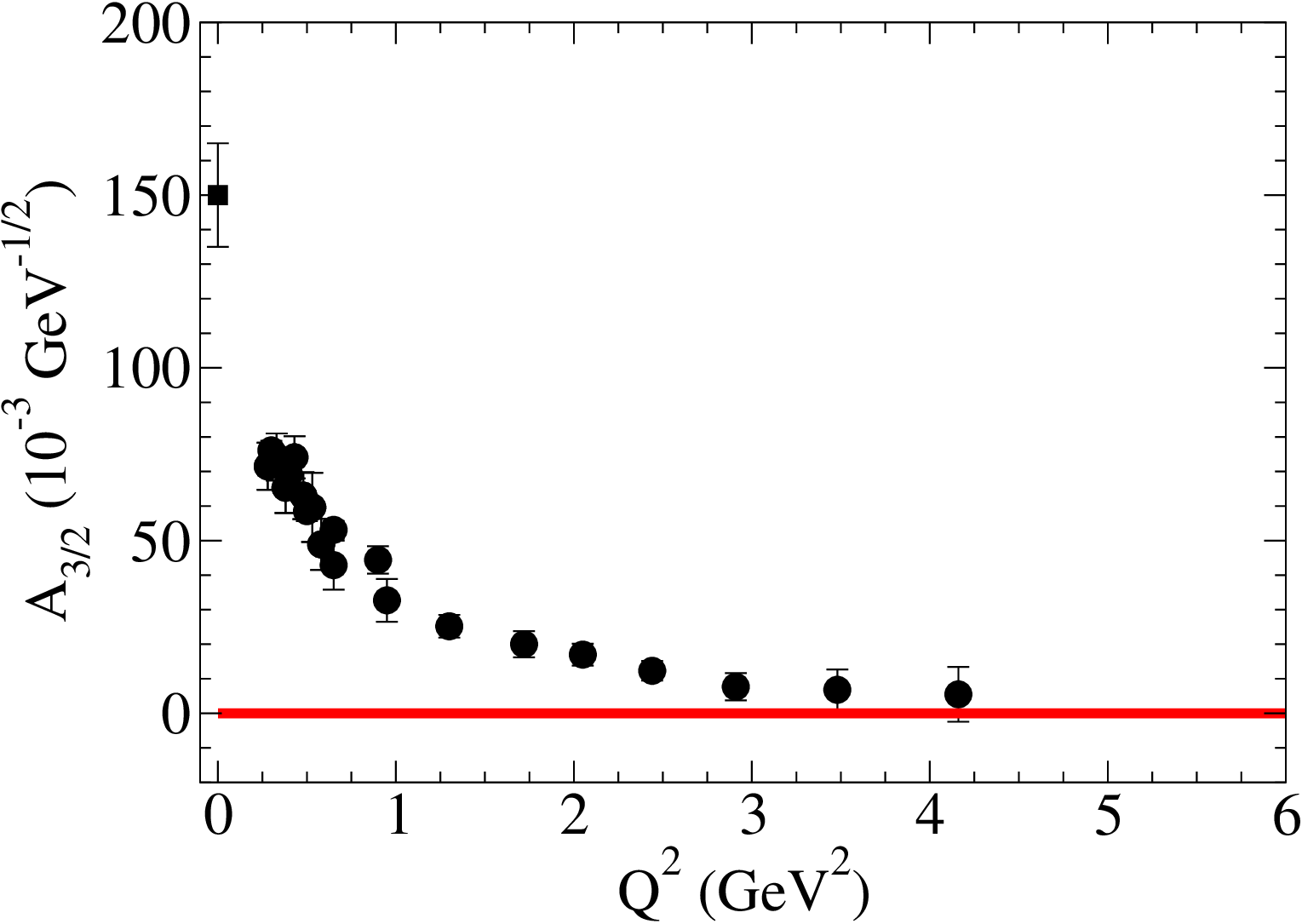}  \hspace{.1cm}
\includegraphics[width=1.45in]{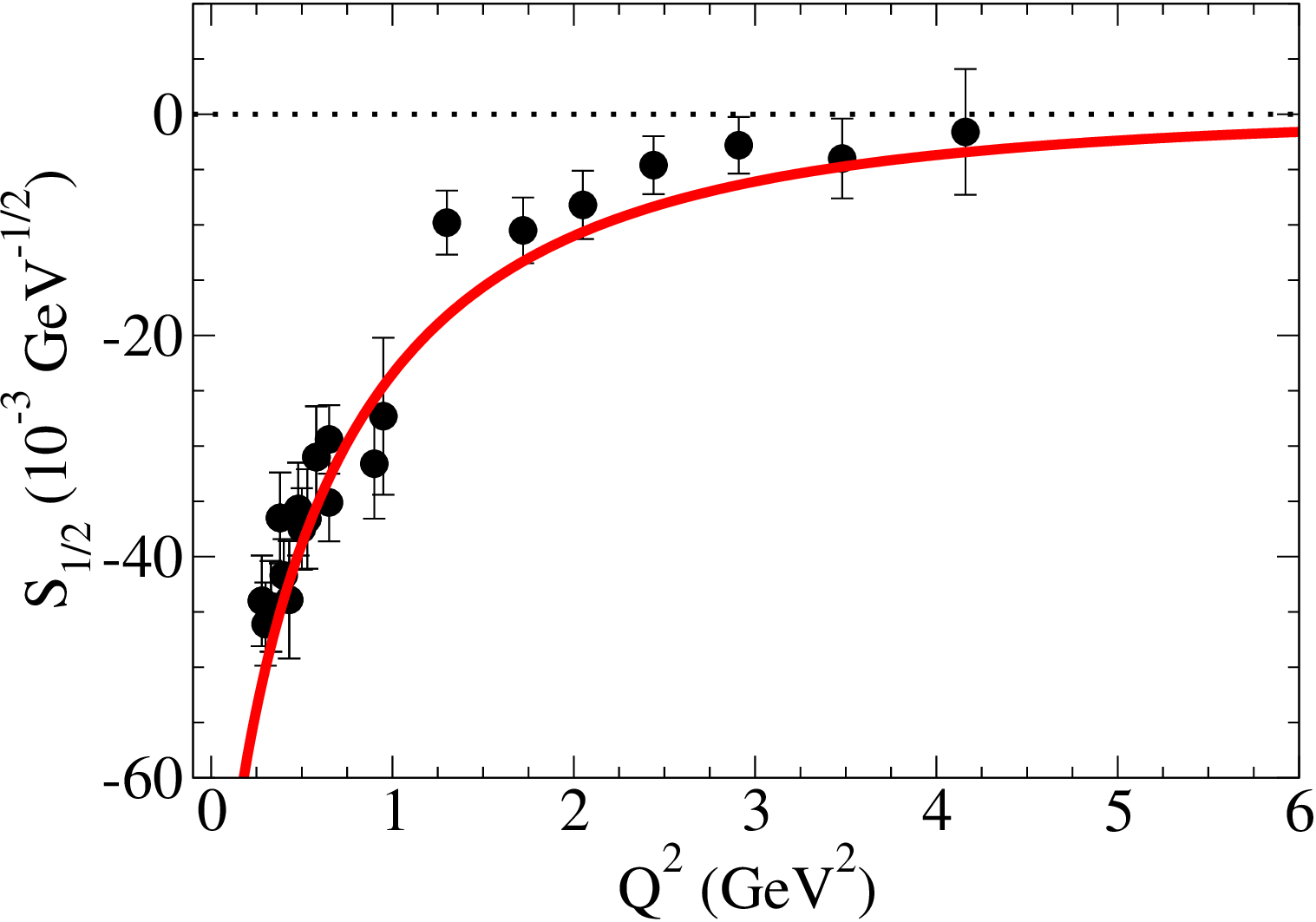} } 
\caption{$\gamma^\ast N \to N(1520)$ amplitudes.
Data from CLAS~\cite{Aznauryan09,Mokeev12,Mokeev16} 
and PDG~\cite{PDG}.}
\label{figN1520-1}       
\end{figure}

\begin{figure}[t]
\mbox{
\includegraphics[width=1.45in]{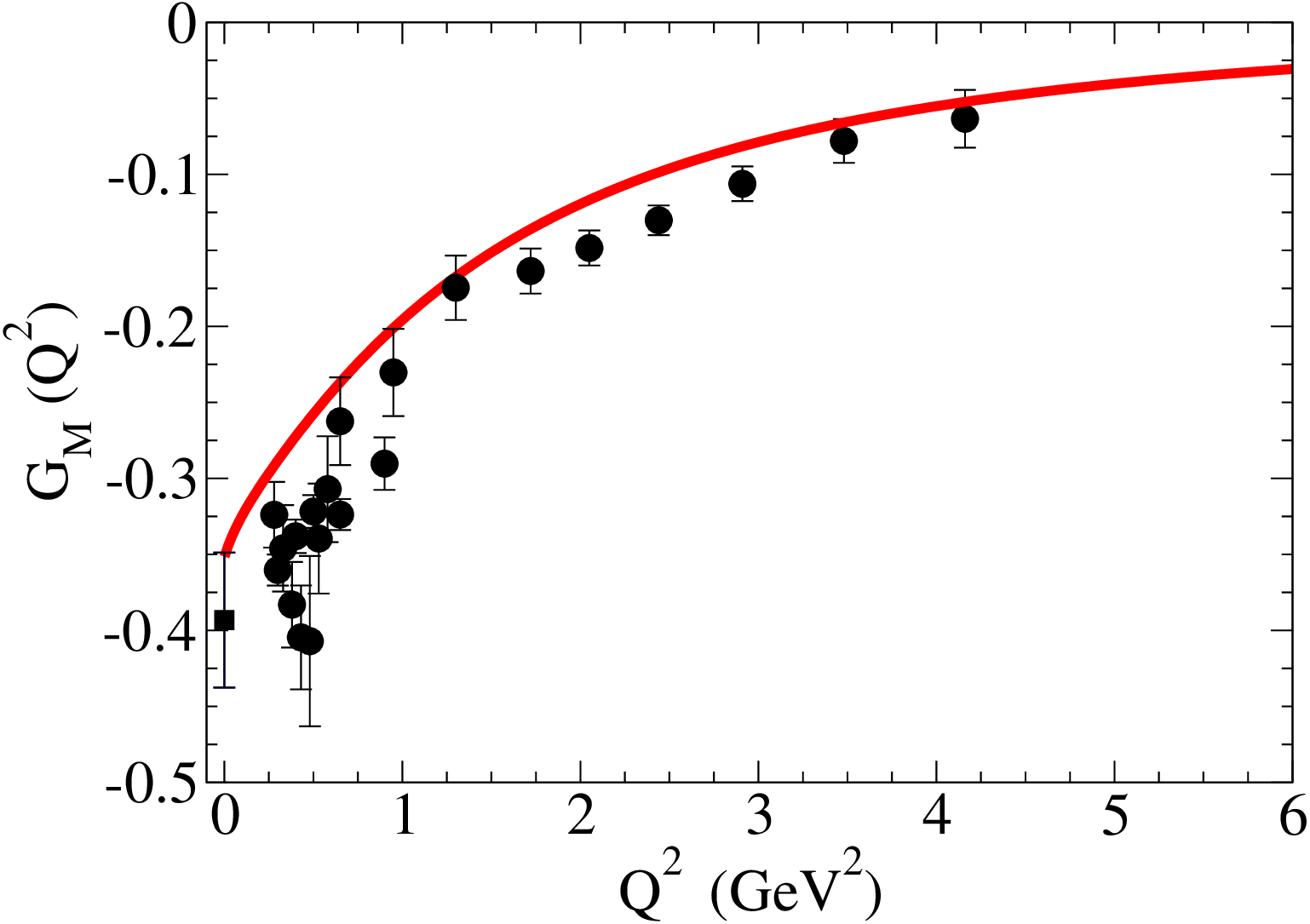}  \hspace{.1cm}
\includegraphics[width=1.45in]{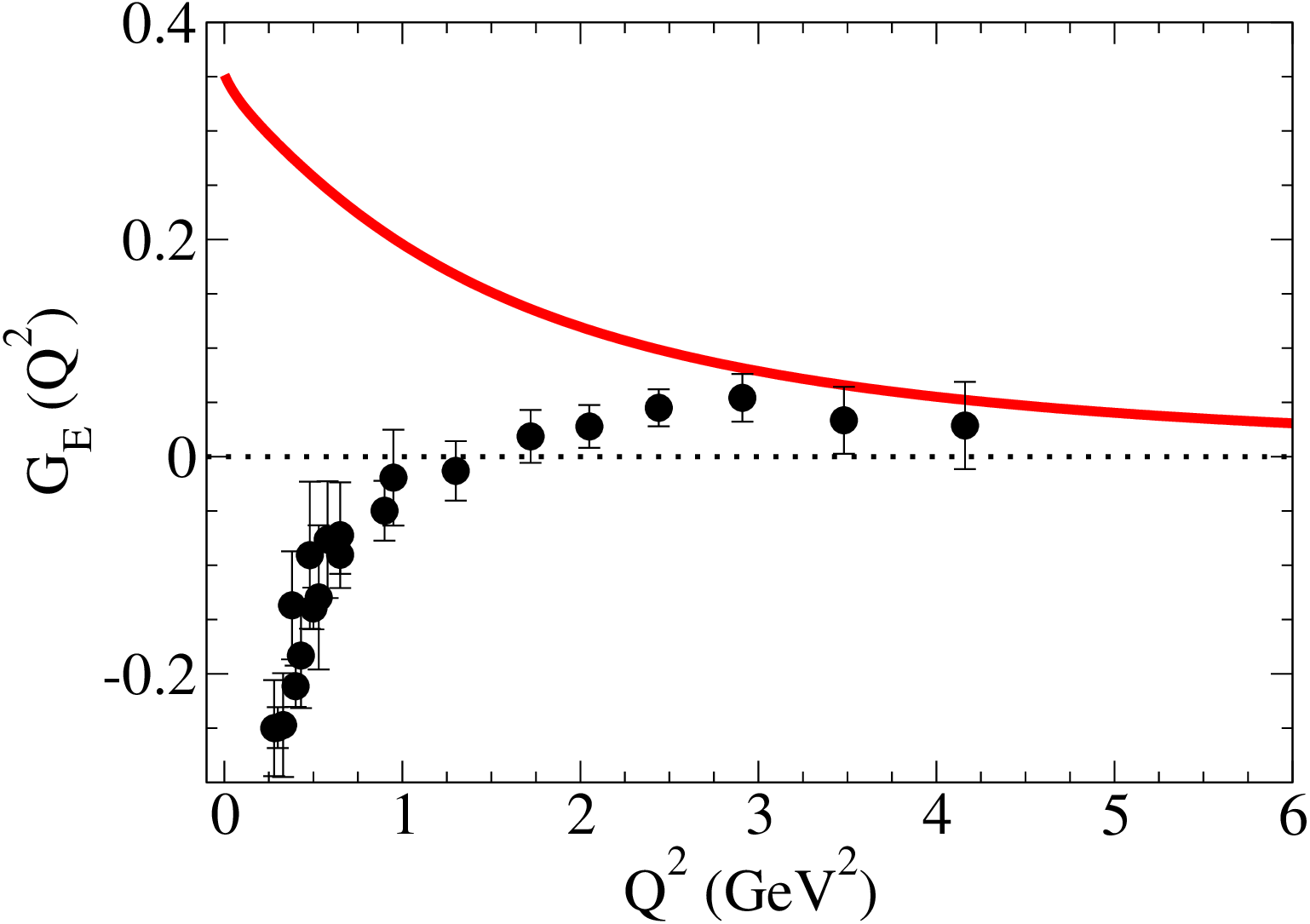}  \hspace{.1cm}
\includegraphics[width=1.45in]{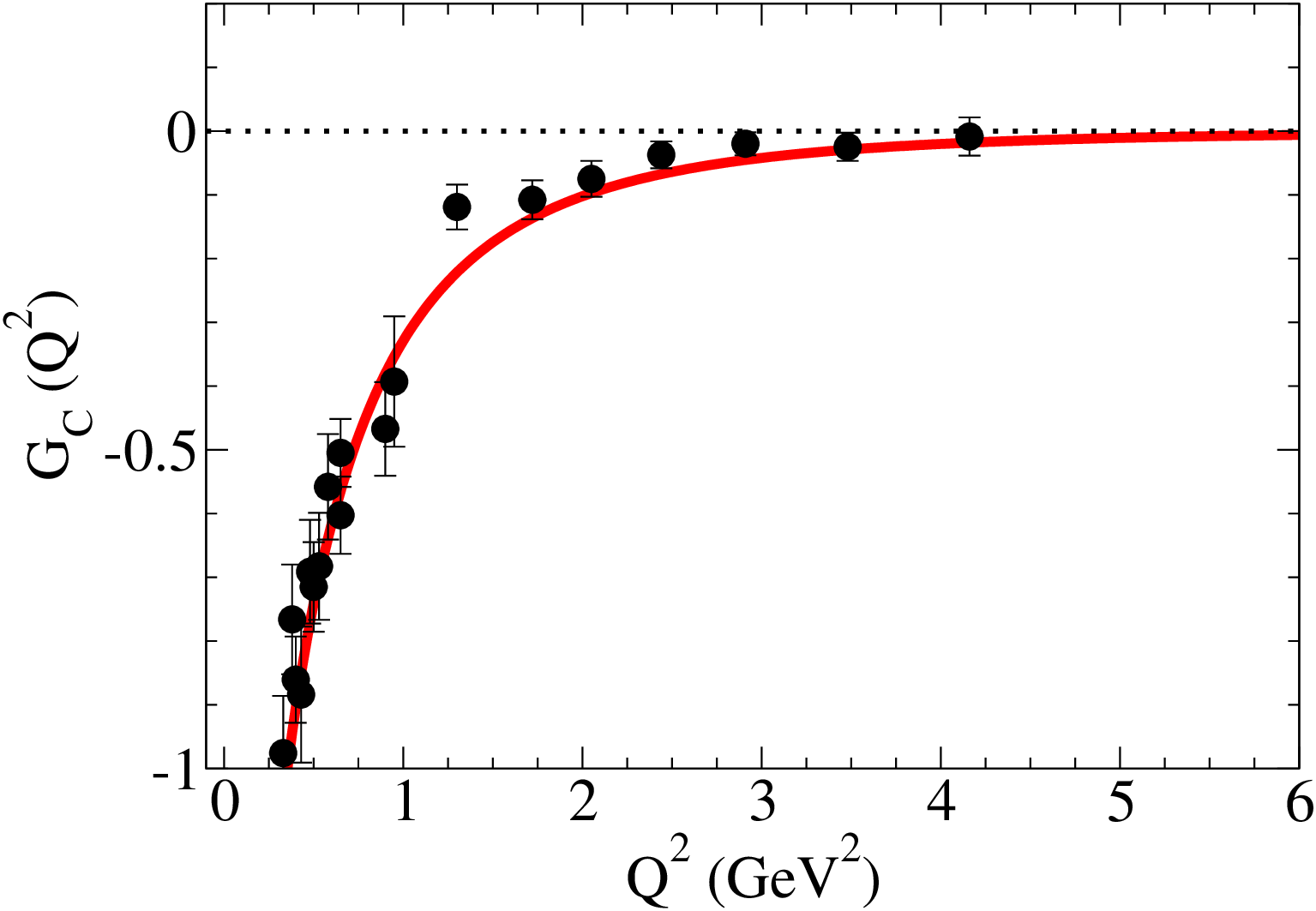}}
\caption{$\gamma^\ast N \to N(1520)$ multipole form factors.
Data from CLAS~\cite{Aznauryan09,Mokeev12,Mokeev16} and PDG~\cite{PDG}.}
\label{figN1520-2}       
\end{figure}

The previous results for the elementary $G_i$ ($i=1,2,3$) can be 
used to calculate the multipole form factors $G_M$, $G_E$ and $G_C$ 
as well as the amplitudes $A_{1/2}$, $A_{3/2}$ and $S_{1/2}$ using 
standard relations, including the explicit dependence on $M_N$ and $M_R$~\cite{N1520,N1520TL}.
The results are presented in Fig.~\ref{figN1520-1} for the helicity 
amplitudes and in Fig.~\ref{figN1520-2} for the transition form factors,
in comparison with the CLAS data~\cite{Aznauryan09,Mokeev12,Mokeev16} 
and the Particle Data Group (PDG) data~\cite{PDG}.

In the recent years there have been a discussion about the 
difference between the analysis of JLab and MAID,
for the helicity amplitudes associated with the $N(1520)$ resonance. 
A detailed discussion can be found in Ref.~\cite{N1520}.
From the presentations of V.~Burkert, V.~Mokeev 
and L.~Tiator~\cite{Burkert-talk,Mokeev-talk,Tiator-talk}, 
we can conclude that this topic is still under discussion.

In Figs.~\ref{figN1520-1} and \ref{figN1520-2} one can see 
that the semirelativistic approximation describe 
very well the empirical data for $Q^2 > 1.5$ GeV$^2$,
with two exceptions: the amplitude $A_{3/2}$ 
and the form factor $G_E$ for small $Q^2$.
Apart these cases one can notice that
the deviation below 1.5 GeV$^2$ is not very significant,
suggesting small meson cloud effects for 
the functions $A_{1/2}$, $S_{1/2}$, $G_M$ and $G_C$.
For that result contributes the relation  
$G_i \propto \sfrac{{\cal I}_R}{|{\bf q}|}$,
which imply that $G_M$ and  $A_{1/2}$ are finite  
in the limit $Q^2=0$~\cite{SRapp}.


The failure of the model for $A_{3/2}$, where 
the model predicts $A_{3/2} \equiv 0$ in contradiction with 
the significant magnitude of the data, 
can be interpreted as a limitation of the model calculation, 
based exclusively on valence quark degrees of freedom~\cite{N1520,N1520TL}.
The magnitude of the  $A_{3/2}$ data can be an indication 
that the amplitude is dominated by meson cloud effects.
This interpretation is corroborated by 
calculations from other authors, which conclude 
that the valence quark contributions can explain only about one third of the observed data.
For a more detailed discussion, check Refs.~\cite{SQTM,SRapp,N1520,N1520TL} 
and references therein.

Concerning to the results for $G_E$, 
the failure of the approximation is related to the result $A_{3/2} =0$,
and the relation $A_{3/2}  \propto (G_M + G_E)$, which implies that $G_M=-G_E$~\cite{N1520,N1520TL}.

The interpretation of the results for $A_{3/2}$ 
as mainly a consequence of the meson cloud effects,
and the conclusion that those effects are not so 
significant for the other functions, in particular $A_{1/2}$,
can be used to estimate the magnitude of the  meson cloud contributions.
The assumption that $|A_{3/2}^{\rm mc}| \gg |A_{1/2}^{\rm mc}|$, where ``mc'' 
label the meson cloud contribution to the amplitudes,
implies that~\cite{SRapp}:
\ba
G_E^{\rm mc} \simeq - \frac{F}{\sqrt{3}}A_{3/2}^{\rm mc}, 
\hspace{1.2cm}
G_M^{\rm mc}  \simeq \frac{1}{3}  G_E^{\rm mc},
\ea 
where $F= \frac{2 M_N}{\sqrt{2 \pi \alpha_{0}}} 
\sqrt{\frac{M_N(M_R + M_N)}{(M_R-M_N) Q_+^2}}$
and $Q_+^2= (M_R + M_N)^2 + Q^2$ ($\alpha_{0}$ is the fine structure constant).
The previous relations allows then the estimation 
of the meson cloud contribution to $G_E$ in terms of the 
empirical value of $A_{3/2}$, and the conclusion 
that the effect is much smaller (one third) in the case of $G_M$.

In Refs.~\cite{N1520,N1520TL} the assumption that $A_{3/2}$ 
is dominated by meson cloud effects was used to derive 
empirical parametrizations for that amplitude.
Those parametrizations can then be used 
in other studies of the $N(1520)$ systems.
An example is the model described in Ref.~\cite{N1520TL}
for the electromagnetic structure of $N(1520)$ in the timelike region.
Another example of the use of those parametrizations is 
discussed in Sect.~\ref{secSQTM}, in the context 
of the single quark transition model.  


\subsection{Summary of the results from the semirelativistic approximation}

The main conclusion of the results from the semirelativistic 
approximation is that one can obtain a very good description 
of the $Q^2 > 1.5$ GeV$^2$ data taking into account the valence quark contributions.
This conclusion is true except for two cases
[$F_2^\ast$ for $N(1535)$ and $A_{3/2}$ for $N(1520)$].

The previous conclusion is impressive, because as mentioned, 
the input in the semirelativistic approximation 
is just the parametrization of the nucleon radial wave function 
and the quark electromagnetic form factors 
(included on the coefficients $j_i^S$ and $j_i^A$).
To summarize, using exclusively the parametrization 
of the nucleon structure, one can estimate, 
with a reasonable precision, 
the transition form factors for the resonances $N(1535)$ and $N(1520)$.

At small $Q^2,$ one can still observe a good description of the data, 
obtaining non-zero results for $G_E$ and $G_M$ at the photon point.
This result is an indication that the meson cloud effects are in general small, 
apart the cases mentioned above.

\subsection{$N(1535)1/2^-$: relation between $A_{1/2}$ and $S_{1/2}$}
\label{secN1535b}

The consequence of the result $F_2^\ast =0$, for $Q^2 > 1.5$ GeV$^2$
is that, the amplitudes $A_{1/2}$ and $S_{1/2}$ are related by~\cite{N1535st}
\ba
S_{1/2} = - \frac{\sqrt{1 + \tau}}{\sqrt{2}} 
\frac{M_R^2- M_N^2}{2 M_R Q} A_{1/2}.
\label{eqScaling}
\ea
The excellent agreement between the $S_{1/2}$ 
data and the r.~h.~s.~was confirmed with great precision 
in the region $Q^2=1.5$--4.2 GeV$^2$ for 
the available data~\cite{N1535st}.
Only future data form JLab-12 GeV upgrade can confirm 
the accuracy for larger values of $Q^2$.

From the theoretical point of view, 
the result (\ref{eqScaling}) can be interpreted 
as a consequence of the cancellation between 
the valence quark contributions and the 
meson cloud contributions for $F_2^\ast$  
at large $Q^2$~\cite{LambdaStar,N1535st}.

Calculations based on the chiral unitary model~\cite{Jido08}, 
which use meson-baryon resonance states
as effective degrees of freedom, are also 
consistent with this interpretation.
The estimates from  the chiral unitary model for $F_2^\ast$,
which can be interpreted as meson cloud contributions, 
generate results comparable in magnitude 
with the estimates from the covariant spectator quark model but 
differ in sign~\cite{LambdaStar,Jido08}.

\section{$N(1440)1/2^+$}
\label{secRoper}

\begin{figure}[t]
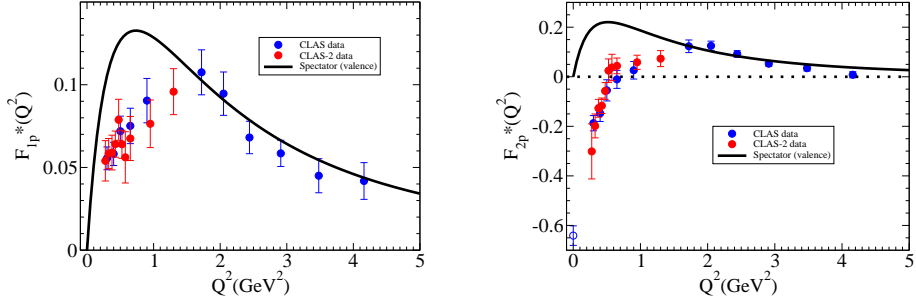

\vspace{.5cm}
\includegraphics[width=5.5cm]{F1_N1440_v3}  \hspace{.8cm}
\includegraphics[width=5.5cm]{F2_N1440_v3} 
\caption{$\gamma^\ast N \to N(1440)$ transition form factors 
for proton target~\cite{Roper,N1710}.
Data from CLAS~\cite{Aznauryan09,Mokeev16} and PDG~\cite{PDG}.}
\label{figRoper1}       
\end{figure}
\begin{figure}[t]
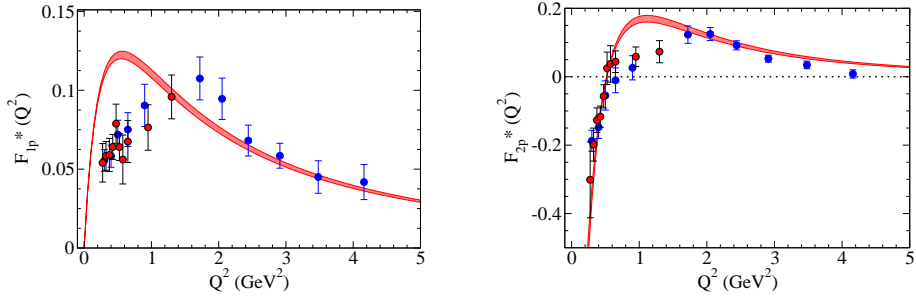

\vspace{.2cm}
\includegraphics[width=5.5cm]{F1R-Tub4}  \hspace{.8cm}
\includegraphics[width=5.5cm]{F2R-Tub4} 
\caption{Holographic estimate of $\gamma^\ast N \to N(1440)$ 
transition form factors for proton target~\cite{RoperHol1}.
The red bands indicate the interval of values 
for the form factors, according to 
the interval of values estimated for the couplings.
Data from CLAS~\cite{Aznauryan09,Mokeev16}.}
\label{figRoper2}       
\end{figure}

The nucleon first radial excitation, the Roper, can be also described by the covariant spectator quark model.
Since $N(1440)1/2^+$ shares with the nucleon the spin-isospin structure,
it differs from the nucleon only by the radial wave function.
One can then derive the form of the Roper radial wave function $\psi_R$
and their relation to the nucleon radial wave function $\psi_N$
imposing the orthogonality between nucleon and Roper states~\cite{Roper,N1710}.
Because the free parameters are fixed by the 
orthogonality condition, the expressions obtained 
for the form factors are true predictions.

The estimates for the form factors are presented in Fig.~\ref{figRoper1}
in comparison with the CLAS data from Refs.~\cite{Aznauryan09,Mokeev16}.
The results for the helicity amplitudes are presented later.
The estimates for the form factors
include only the contributions from the valence quarks.
In the figure, one can notice that the theoretical results
are very close to the empirical data for $Q^2> 1.5$ GeV$^2$,
supporting the assumption that the  $N(1440)1/2^+$
is in fact the first radial excitation of the nucleon~\cite{Aznauryan07}.
Below 1.5 GeV$^2$, the deviation from the data may be 
interpreted as a manifestation of the meson cloud.
In these conditions, one can use our estimate 
of the quark core contribution to estimate the meson 
cloud effect from the CLAS data~\cite{RoperAIP}.

Recently, the $\gamma^\ast N \to N(1440)$ transition form factors
have been estimated using the formalism 
of Holographic QCD~\cite{Brodsky15}.
In this description, the valence quark contributions 
are interpreted as the contributions from the leading order 
Fock state associated with a three-valence quark system~\cite{RoperHol1}.
The 3 independent couplings are first adjusted 
by the nucleon elastic form factor data for large $Q^2$ 
(small meson cloud effects).
The intervals of values obtained for the parameters 
are used to estimate the $\gamma^\ast N \to N(1440)$ transition form factors,
presented in  in Fig.~\ref{figRoper2} (see red band).
In the figure we can notice that 
the holographic estimate describes also very well 
the large-$Q^2$ region.


\begin{figure}[t]
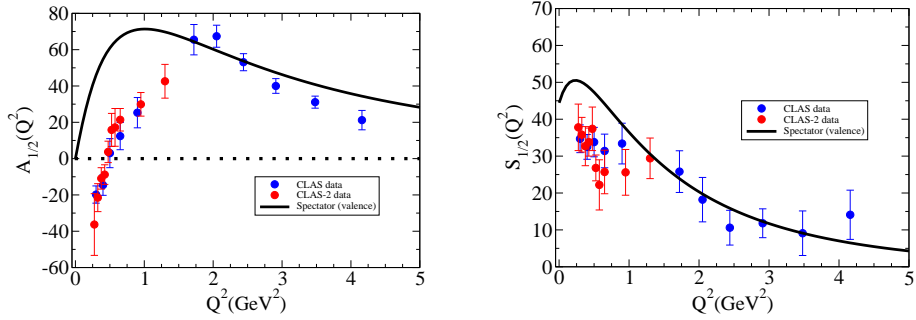

\vspace{.5cm}
\includegraphics[width=5.5cm]{A12_N1440_v3}  \hspace{.8cm}
\includegraphics[width=5.5cm]{S12_N1440_v3} 
\caption{$\gamma^\ast N \to N(1440)$ helicity amplitudes 
for proton target~\cite{Roper,N1710}.
Data from CLAS~\cite{Aznauryan09,Mokeev16} and PDG~\cite{PDG}.}
\label{figRoper1-amps}       
\end{figure}
\begin{figure}[t]
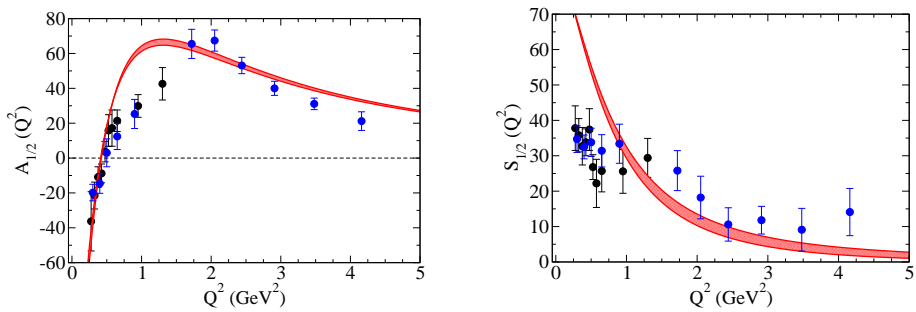

\includegraphics[width=5.5cm]{A12-N1440-Hol1}  \hspace{.8cm}
\includegraphics[width=5.5cm]{S12-N1440-Hol1} 
\caption{Holographic estimate of $\gamma^\ast N \to N(1440)$ 
helicity amplitudes for proton target~\cite{RoperHol1}.
The red bands indicate the interval of values 
for the form factors, according to 
the intervals of values estimated for the couplings.
Data from CLAS~\cite{Aznauryan09,Mokeev16}.}
\label{figRoper2-amps}       
\end{figure}

The results from Fig.~\ref{figRoper2} for the Pauli form factor $F_{2p}^\ast$
are surprising, because they show an excellent agreement 
with the low-$Q^2$ data.
This result suggests that the meson cloud contributions to 
$F_{2p}^\ast$ may be very small contrarily 
to what it is usually expected for the transition form factors.

The calibration of the bare couplings determined in Ref.~\cite{RoperHol1}
can also be used to derive analytic parametrization 
to the $\gamma^\ast N \to N(1440)$ transition 
form factors $F_{1p}^\ast$ and $F_{2p}^\ast$~\cite{RoperHol2}.

Overall, one can conclude that Holographic QCD is a very promising 
method to study  the $\gamma^\ast N \to N^\ast$ transitions, 
since it provides a useful tool to estimate the valence quark effects at small $Q^2$.

The results for helicity amplitudes, transverse ($A_{1/2}$)
and longitudinal ($S_{1/2}$) corresponding to the 
transition form factors from Figs.~\ref{figRoper1} and \ref{figRoper2}
are presented in the Figs.~\ref{figRoper1-amps} and \ref{figRoper2-amps},
for the covariant spectator quark model and the holographic model, respectively.
In general, one can observe a good agreement with the data for $Q^2 > 1.5$ GeV$^2$, 
as expected from the analysis of the transition form factors.
For the holographic model, in particular, 
one can observe at small $Q^2$, the underestimation of the $S_{1/2}$ data.
This result is mainly the consequence of the model results for $F_{1p}^\ast$, 
where there is an overestimation of the data 
at small $Q^2$ (see Fig.~\ref{figRoper2}).
This happens because $S_{1/2} \propto (F_{1p}^\ast  - \sfrac{Q^2}{(M_R+M)^2} F_{2p}^\ast)$,
which implies a reduction of the effect of $F_{2p}^\ast$ 
in this amplitude for small values of $Q^2$. 
A more detailed discussion of the holographic estimates of the helicity amplitudes 
at small $Q^2$ can be found in Ref.~\cite{RoperHol2}.


\section{$\Delta(1232)3/2^+$}
\label{secDelta}

The $\gamma^\ast N \to \Delta(1232)$ is characterized by 
the 3 mutipole form factors: the magnetic dipole ($G_M^\ast$),
the electric quadrupole ($G_E^\ast$) and the Coulomb quadrupole 
($G_C^\ast$)~\cite{NDelta,NDeltaD,Jones73}.
According with the $SU(6)$ symmetry,
the $\gamma^\ast N \to \Delta(1232)$ transition
is predominantly a magnetic transition, 
as a consequence of a spin-flip of a 
quark on the nucleon to create the $\Delta(1232)$ (spin 3/2)~\cite{JDiaz07}.
As a result, $G_M^\ast$ is the dominant form factor in the transition.

There are, however, also contributions from the quadrupole form factors, 
$G_E^\ast$ and $G_C^\ast$.
The contributions of these form factors for the transition cross-section
are determined by the functions $G_E^\ast$ and 
$\frac{|{\bf q}|}{2 M_\Delta}G_C^\ast$,
and have therefore a small impact
 ($|{\bf q}|$ is the photon three-momentum and $M_\Delta$ is the 
$\Delta(1232)$ mass)~\cite{NDeltaD,LatticeD,Siegert-ND,RSM-Siegert,GlobalFit,Siegert1}.

\begin{figure}
\vspace{.5cm}
\begin{center}
\includegraphics[width=7cm]{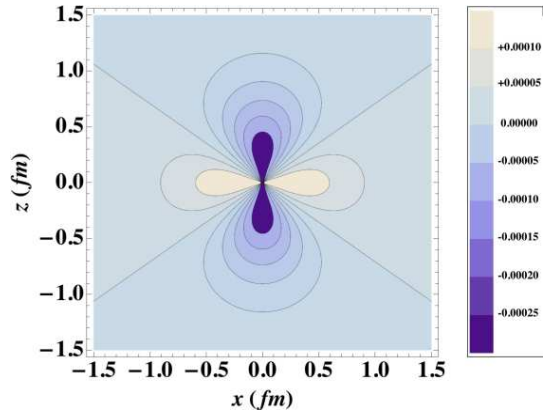}  
\end{center}
\caption{Visualization of the $\Delta^+(1232)$ deformation. 
The density is defined for the maximum spin projection $S_z= + \frac{3}{2}$~\cite{Deformation}.
The lither regions correspond to the larger values
(the scale is defined on the right side).}
\label{figDeltaDef}       
\end{figure}

The small but nonzero contributions for $G_E^\ast$ and $\frac{|{\bf q}|}{2 M_\Delta}G_C^\ast$ 
can be interpreted as an indication of the deformation 
from the $\Delta(1232)$~\cite{Deformation,Becchi65,Buchmann00b,Alexandrou09}.
The present results for the quadrupole form factors are compatible 
with an oblate shape for the $\Delta^+(1232)$~\cite{Deformation,Buchmann00b}.
The estimate of the valence quark contributions for the $\Delta^+(1232)$
spacial density is presented in Fig.~\ref{figDeltaDef}.
In the figure, one can confirm the expected deformation along 
the horizontal axis for the $\Delta^+(1232)$
(the vertical axis is defined by the spin projection: 
$S_z= + \sfrac{3}{2}$)~\cite{Deformation}.

In the following, we discuss the results for 
the $\gamma^\ast N \to \Delta(1232)$ form factors 
based on the calculations from the covariant spectator quark model.

\subsection{Magnetic dipole form factor}

The magnetic dipole form factor is in general dominated 
by the valence quark contributions, 
consistently with estimates based on $SU(6)$, 
where the quarks are described by $S$-states~\cite{JDiaz07,Pascalutsa07b}.
It is well known, however, that models based 
on valence quarks underestimate the experimental data 
in about 40\% at low $Q^2$~\cite{JDiaz07,SatoLee,Kamalov,Pascalutsa07b}.
This subject was also discussed in the presentations 
from H.~Kamano and V.~Burkert~\cite{Kamano-talk,Burkert-talk}.

The missing strength in the form factor $G_M^\ast$ 
is in general interpreted as the consequence of the meson cloud,
dominated the lightest meson, the pion, 
which is not not taken into account 
in quark model calculations~\cite{NDelta,JDiaz07,SatoLee,Kamalov}.
In the case of the covariant spectator quark model, this underestimation 
can be naturally explained 
when we consider a simple model where the nucleon and the $\Delta(1232)$
are described by $S$-states for the quark-diquark system.
In these conditions the quadrupole form factors have no contributions 
($G_E^\ast = G_C^\ast \equiv 0$) and only the magnetic dipole form 
factor have non-zero results~\cite{NDelta}.
The magnetic dipole form factor can then be 
calculated using the a simple expression~\cite{NDelta,NDeltaD}: 
\ba
G_M^\ast (Q^2)= \frac{4}{3 \sqrt{3}} \frac{M_N}{M_\Delta + M_N}
\left[ f_{1-}  +  \frac{M_\Delta + M_N}{2 M_N} f_{1-}  \right] 
\int_k \psi_\Delta \psi_N 
\ea 
where $f_{1-}$ and $f_{2-}$ are the quark Dirac and Pauli 
isovector form factors and 
$\psi_N$, $\psi_\Delta$ are the nucleon and $\Delta(1232)$ radial 
wave functions.
Based on the previous relation, one can conclude, 
using the normalization conditions, the  
Cauchy-Schwartz  inequality, $\int_k\psi_\Delta \psi_N  \le 1$,
and the values of the quark anomalous magnetic moments, that 
$G_M^\ast \le 2.07$ for $Q^2=0$~\cite{NDelta,OctetDecuplet}.

Since the experimental value is $G_M^\ast(0) \simeq 3.02$~\cite{PDG},
one can conclude that near $Q^2=0$, 
the model underestimates the data in about 37\%.
Note that this estimate provide only an upper limit,
and that in practice,
one can have even larger underestimations~\cite{NDelta,OctetDecuplet}.

From the previous discussion, we can conclude that 
the covariant spectator quark model provides a natural explanation for 
the underestimation of the data at low $Q^2$, when we consider only the  
valence quark degrees of freedom.
In order to explain the missing strength, one needs to 
take into account explicit contributions of the pion cloud effects,
as concluded from the use of 
dynamical baryon-meson reaction 
models~\cite{Burkert04,NDelta,JDiaz07,SatoLee,Kamalov}.

Before explaining how one can parametrize the pion cloud effects,
one needs to discuss how we can parametrize the  
of the nucleon and the $\Delta(1232)$ wave functions.
As discussed in Sect.~\ref{secCSQM}, the structure of the nucleon 
can be described within the covariant spectator quark model, considering 
an $SU(6)$ structure for the $S$-state wave function, and a parametrization 
for the quark current (\ref{eqjq})~\cite{Nucleon}.
As for the nucleon, we consider also an $S$-state structure 
associated with a radial wave function $\psi_\Delta$~\cite{NDelta,NDeltaD,LatticeD}.
The question is, how to determine the function $\psi_\Delta$, 
since, contrary to the nucleon elastic form factors, 
the radial wave function cannot be adjusted directly to the empirical data,
because the data is strongly contaminated by pion cloud effects.
 
We are then left with two options:
i) calibrate the data by some estimate from the valence quark 
core contributions to the transition form factors;
ii) calibrate the model by lattice QCD simulations 
for large pion masses, where the meson cloud effects are suppressed.

\begin{figure}[t]
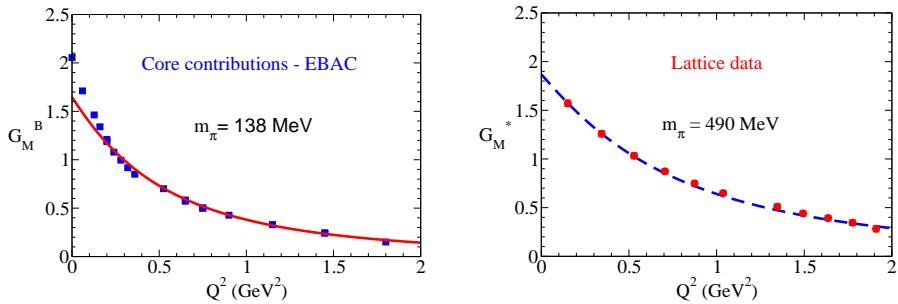

\vspace{.5cm}
\centerline{
\mbox{
\includegraphics[width=5.5cm]{GMS138b}  \hspace{.5cm}
\includegraphics[width=5.5cm]{GM490} }}
\caption{{\bf At the right:} Comparison with EBAC estimate of bare core~\cite{JDiaz07}.
{\bf At the left:} Extrapolation to the lattice QCD regime 
with $m_\pi= 490$ MeV. Lattice data from Ref.~\cite{Alexandrou08}.}
\label{figGM1}       
\end{figure}

The first option can be implemented using the estimate 
of the quark core contributions performed with the assistance 
of the Sato-Lee/EBAC model, nowadays known as 
Argonne-Osaka model~\cite{EBACreview,JDiaz07,EBAC,Kamano-talk}.
The second option requires an intermediate step, 
the extension of the covariant spectator quark model from the physical regime 
to the lattice QCD regime. 
This extension can be performed taking advantage 
of the definition of the quark currents in terms 
of the hadron masses (vector mesons and nucleon mass) 
and also the convenient definition of the radial 
wave functions in terms of the mass of the baryons under study.
The discussion about the extension of the 
covariant spectator quark model for lattice QCD 
can be found in Refs.~\cite{Omega,LatticeD,Octet,Omega2,Lattice}.   
 
The results from the covariant spectator quark model for $G_M^\ast$ 
are presented in Fig.~\ref{figGM1}.
The parameters of the model are adjusted to the 
EBAC estimate of the bare core in Ref.~\cite{NDeltaD}.
The results are presented in the left panel, 
where $G_M^\ast$ is relabeled as $G_M^B$ 
($B$ holds for bare), since no pion cloud contributions are 
taken into account.
The label  $m_\pi= 138$ MeV is included to remind 
that the calculation is performed at the physical point.
The model was later extended to the lattice QCD regime
using the same parametrization for $\psi_\Delta$~\cite{LatticeD}.
The extension of the model gives as accurate description 
of the lattice QCD data for $m_\pi= 411$, 490 and 563 MeV~\cite{Alexandrou08}.
The results for  $m_\pi= 490$ MeV are presented in the left panel
of Fig.~\ref{figGM1}.

Since the parametrization for $\psi_\Delta$ 
gives a simultaneously good description of the 
EBAC estimate of the bare core contribution, 
and of the lattice data for $m_\pi > 0.4$ GeV, 
a region where we expect very small pion cloud effects,
we can conclude that we have a consistent description 
of the physics associated with the valence quark degrees of freedom.

\subsection*{How to simulate the pion cloud ? }

As discussed above,
if we want to describe the $\gamma^\ast N \to \Delta(1232)$ 
magnetic dipole form factor in all range of $Q^2$, 
one needs to take into account the mechanism of the pion cloud.

\begin{figure}[t]
\vspace{.8cm}
\centerline{
\mbox{
\includegraphics[width=7cm]{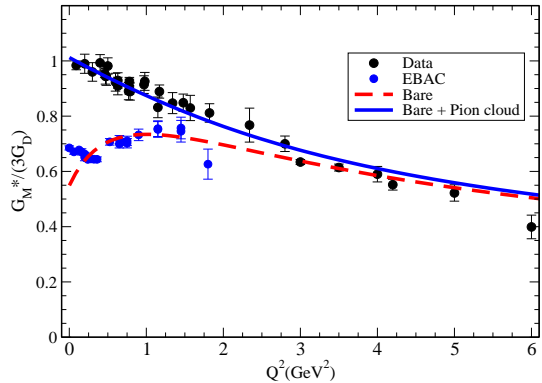}}}
\caption{Combination of valence (Bare) 
and pion cloud contributions for $G_M^\ast$~\cite{LatticeD}.
The valence quark contributions are the same as in Fig.~\ref{figGM1}. 
The pion cloud contribution is estimated by Eq.~(\ref{eqGMpi}).
The blue bullets represent the EBAC estimate of the quark core~\cite{JDiaz07},
as before.}
\label{figGM2}       
\end{figure}

In our first attempt, we considered a simple phenomenological 
parametrization of the pion cloud contribution to $G_M^\ast$, $G_M^\pi$,
in the form~\cite{NDelta,NDeltaD,LatticeD}
\ba
\frac{G_M^\pi}{3 G_D} = \lambda_\pi 
\left( \frac{\Lambda_\pi^2}{\Lambda_\pi^2 + Q^2}\right)^2,
\label{eqGMpi}
\ea
where $G_D = 1/(1 + Q^2/0.71)^2$ is the dipole form factor,
$\lambda_\pi$ define the strength 
of the contribution, and $\Lambda_\pi^2$ is a cutoff that define
the falloff.
The best description of the data is obtained 
for  $\lambda_\pi= 0.441$ and $\Lambda_\pi^2=1.53$ GeV$^2$~\cite{NDeltaD}. 
 
The representation (\ref{eqGMpi}) is motivated 
by the usual representation of $G_M^\ast$, 
normalized by the factor $3 G_D$, 
and also for the expected falloff from pertubative QCD (pQCD).
Based on pQCD arguments one can conclude 
that a contribution associated with the pion 
on a three-quark system, and represented by a $q \bar q$
state is described by a function $\propto 1/Q^{2(3 + 2-1)} = 1/Q^8$,
where the extra 2 is the consequence of the additional $q \bar q$ state~\cite{Carlson}.

\begin{figure}[t]
\vspace{.8cm}
\centerline{
\mbox{\includegraphics[width=8cm]{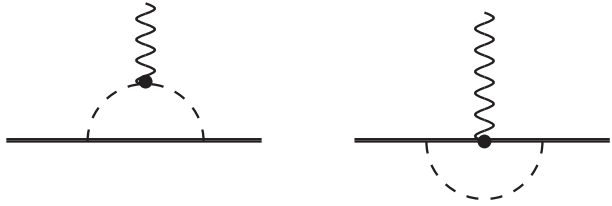}}}
\caption{Diagrammatic representation of first 
contributions to the pion cloud contributions.
The first diagram represents the photon coupling 
with the pion (includes function $F_\pi$).
The second diagram simulates the photon interaction 
with intermediate baryon states.}
\label{figPionCloud}       
\end{figure}

The result of the parametrization $G_M^\pi$, 
combined with the valence quark contributions discussed 
in the previous section (see Fig.~\ref{figGM1}) 
is presented in Fig.~\ref{figGM2}, up to 6 GeV$^2$.
In the figure, one can see that 
the bare contributions give a good description of the data 
when  $Q^2 > 3$ GeV$^2$ (small pion cloud).

Once tested the parametrization to the pion cloud component,
one can use it in calculations associated with the  
 $\gamma^\ast N \to \Delta(1232)$  transition.
The parametrization was used, in particular 
in the calculation of the octet to 
decuplet transition form factors~\cite{OctetDecuplet,OctetDecuplet2},
and also in the calculation of the 
$\gamma^\ast N \to \Delta(1600)$ transition form factors~\cite{Delta1600}.
An extension of Eq.~(\ref{eqGMpi}) with minor modifications 
was used in the calculation of the $\gamma^\ast N \to \Delta(1232)$ 
transition in the timelike regime~\cite{NDeltaTL}.


In the more recent studies of the  $\gamma^\ast N \to \Delta(1232)$  transition,
we considered an improved form of the parametrization (\ref{eqGMpi}).
One considers in particular the form~\cite{NDeltaTL2}
\ba
G_M^\pi(q^2)= 3 \frac{\lambda_\pi}{2} F_\pi (q^2) 
\left( \frac{\Lambda_\pi^2}{\Lambda_\pi^2 - q^2}\right)^2 +
  3 \frac{\lambda_\pi}{2}  
\left[
\frac{\Lambda_D^4}{(\Lambda_D^2-q^2)^2 +  \Lambda_D^2\Gamma_D^2(q^2) }
\right]^2,
\label{eqGMpi2}
\ea
where $F_\pi(q^2)$ is the empirical pion electromagnetic form factor,
$\Lambda_D^2=0.9$ GeV$^2$ is the nucleon dipole cutoff,
and $\Gamma_D$ is a phenomenological width.
In this case the functions are represented in terms of $q^2= -Q^2$, 
in order to facilitate the discussion in the timelike regime.
The new parametrization improves the previous one, 
because it  clearly separates
the contributions from the photon coupling with the pion 
from the photon coupling with intermediate baryon states 
(see Fig.~\ref{figPionCloud}).


The motivation for the use of the parametrization (\ref{eqGMpi2}) 
is based on the diagrammatic representation of Fig.~\ref{figPionCloud},
and in the results of the study of the octet to decuplet transition 
from Ref.~\cite{OctetDecuplet2}.
In that work a microscopic meson cloud contribution 
based on the cloudy bag model \cite{CBM} was used
in combination with the covariant spectator quark model for the quark core.
It was found that in the case of the 
$\gamma^\ast N \to \Delta(1232)$  transition
each diagram contribute with about 50\% to the pion cloud effect.
 
In the new representation only a part (50\%) of the contribution 
is then linked with the photon coupling with the pion, 
as expected in a realistic description.
The second term, which describes the coupling with intermediate baryons
is now represented phenomenologically, 
using an effective generalization of $G_D^2$ to the timelike region, 
where the pole $q^2= \Lambda_D^2$ is regularized~\cite{NDeltaTL,NDeltaTL2}.

The present representation of $G_M^\pi$ 
is particularly useful for studies in the timelike region,
in particular to the study of the $\Delta(1232)$ Dalitz decay:
$\Delta \to \gamma^\ast  N \to e^+ e^- N$, where the final state 
has a dilepton pair~\cite{NDeltaTL2,HADES10}.
These processes have been studied at HADES~\cite{HADES10,MesonBeams,Weil12,HADES17}.
This topic was discussed also in the presentation of B.~Ramstein~\cite{Ramstein-talk}.

\begin{figure}[t] 
\vspace{.5cm}
\includegraphics[width=5.5cm]{GMTL_W-mod4}  \hspace{.4cm}
\includegraphics[width=6.1cm]{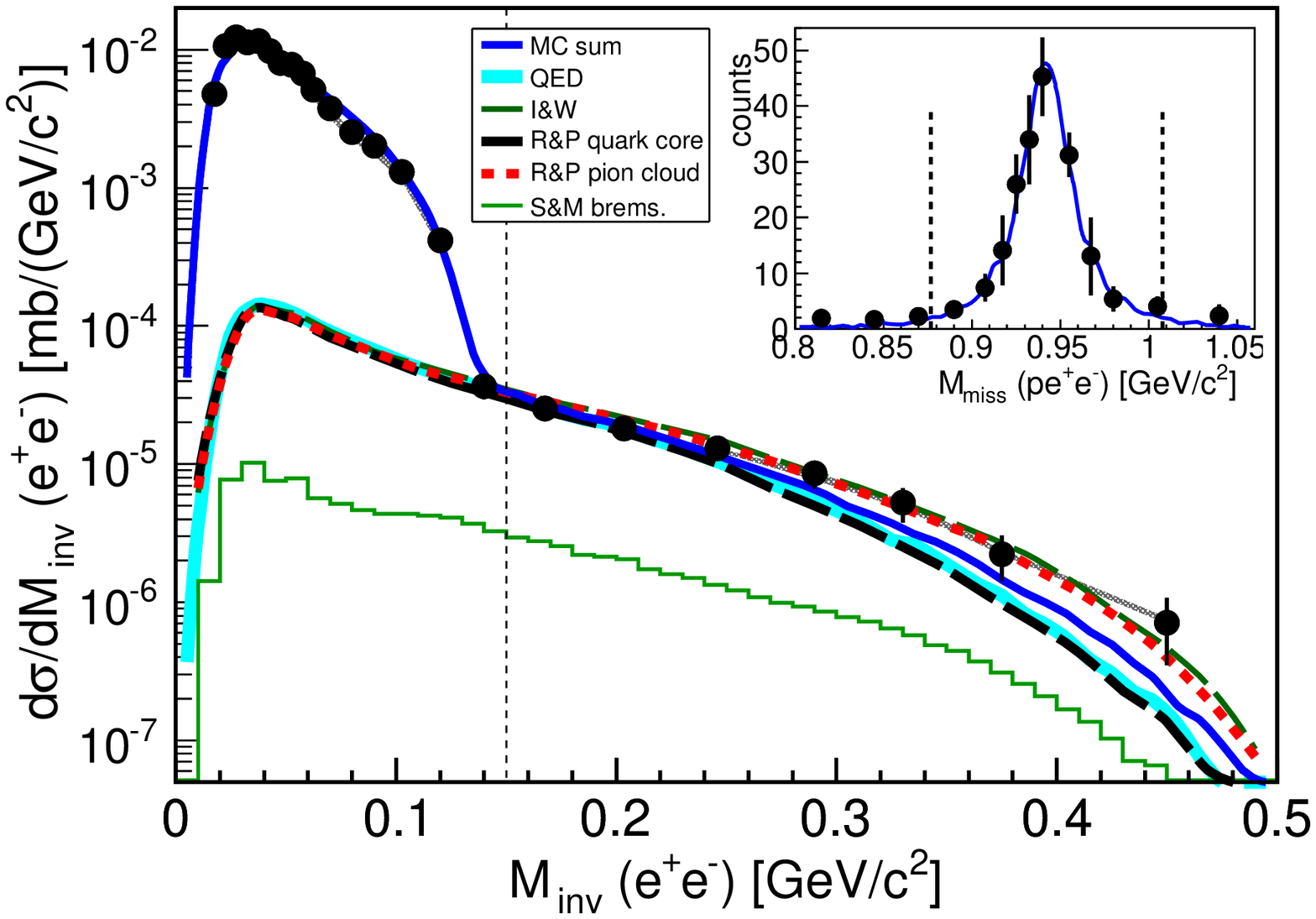} 
\caption{{\bf At the right:} 
Calculation of $|G_M^\ast|$ in timelike region in terms of $W$~\cite{NDeltaTL2}.
{\bf At the left:} $\Delta(1232)$ Dalitz decay cross-sections from HADES~\cite{HADES17}. 
See discussion in the main text.}
\label{figGMTL}       
\end{figure}

In the timelike region, one can calculate 
the $G_M^\ast$ form factor, which is complex, 
in terms of the running mass $W$ that can differ from the mass of the pole $M_\Delta$.
The results of $|G_M^\ast|$ for different values of $W$
are presented in the left panel of Fig.~\ref{figGMTL}.
For kinematic reasons the functions are limited by 
\mbox{$q^2 \le (W-M)^2$}~\cite{NDeltaTL,NDeltaTL2}.
The model for $|G_M^\ast|$ was used to estimate 
the $\Delta(1232)$ Dalitz cross-sections and it was  
compared with the results from HADES~\cite{HADES17}.
The results are presented in the right panel of Fig.~\ref{figGMTL}.
The covariant spectator quark model (dashed black line) 
gives a very good description of the data below 0.15 GeV/$c^2$
but underestimates the data for large values of 
the $e^+e^-$ invariant mass.
The description of the higher region is achieved
when we take into account 
only the pion cloud contributions (dotted red line)~\cite{HADES17}.

It is expected that in a near future the experimental  
study of the Dalitz decays can be extended to the $N(1520)$~\cite{N1520TL}.


\subsection{Quadrupole form factors}

\begin{figure}[t]
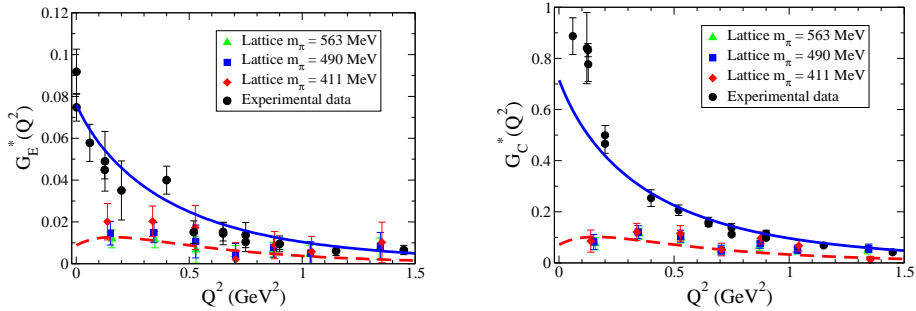

\vspace{.5cm}
\includegraphics[width=5.5cm]{GEexp2}  \hspace{.8cm}
\includegraphics[width=5.5cm]{GCexp2} 
\caption{{\bf At the right:} Results for $G_E^\ast$.
{\bf At the left:} Results for $G_C^\ast$.
In both cases the dashed-line represent the bare contribution 
(at the physical point) and the solid-line the 
combination of the bare and pion cloud.
Lattice QCD data from Ref.~\cite{Alexandrou08}.
The description of the data can be found in Ref.~\cite{LatticeD}.}
\label{figGEGC1}       
\end{figure}

Also for the quadrupole form factors we 
can consider a decomposition between the bare and pion cloud contributions.
The main difference is that in this case 
the bare contributions are very small, 
because those contributions are the consequence 
of small $D$-state mixtures in the $\Delta(1232)$ wave function.
There are two $D$-state to be considered 
in the  $\Delta(1232)$: one where the sum of the quark spin (core spin)
is 1/2; another where the core spin is 3/2.
The $D$-state associated with the core spin 3/2 
gives the dominant contribution to $G_E^\ast$.
The  $D$-state associated with the core spin 1/2 
gives the dominant contribution to $G_C^\ast$~\cite{NDeltaD}.
An accurate description of the lattice QCD data 
favors a mixture of 0.72\% for both $D$-states~\cite{LatticeD}.
The corresponding results for the form factors at the physical limit  
($m_\pi= 138$ MeV) are presented in Fig.~\ref{figGEGC1} by the dashed-line.
In the same figure we present also the 
lattice QCD data from Ref.~\cite{Alexandrou08}
for $m_\pi=411$, 490 and 591 MeV.
From the figure it is clear that the 
estimate of the valence quark contributions 
and the lattice QCD data is way below the physical data, 
represented by the dark bullets.
Near $Q^2=0$ the valence quark contributions 
explain only about 10--20\% of the measured data~\cite{LatticeD,Siegert-ND}.


The predictions approach the experimental data only 
when we include the contributions of the  pion cloud. 
In the literature, there are simple parametrizations 
of the pion cloud contributions to the form factors 
$G_E^\ast$ and $G_C^\ast$, derived 
in the large $N_c$ limit~\cite{Pascalutsa07a,Buchmann02}.
These relations estimate the pion cloud contributions 
for the quadrupole form factors based on parametrizations 
of the neutron electric form factor $G_{En}$.
The result of the combination of valence quark 
and pion cloud contributions is presented 
in Fig.~\ref{figGEGC1} by the solid line.
From the final result, we can conclude that 
the pion cloud effects dominate both form factors.

It is important to note that the estimates of pion cloud contributions, 
are valid only for small $Q^2$, because they are derived 
from the low $Q^2$ expansion $G_{En} \simeq - \sfrac{1}{6}r_n^2Q^2$,
where $r_n^2$ is the neutron electric 
square radius~\cite{Pascalutsa07a}.
It is expected that those contributions are suppressed
for very large $Q^2$ comparative to 
the valence quark contributions, 
according with the pQCD falloffs 
$G_E^\ast \propto 1/Q^4$ and  
$G_C^\ast \propto 1/Q^6$~\cite{Siegert-ND,RSM-Siegert,Carlson}.

From the figure, we can also conclude that there is an excellent 
agreement between model and data when we combine the two effects.
This agreement is impressive because there is no direct fit to the data. 
The valence quark contribution is estimated exclusively 
by the lattice QCD data, and the pion cloud contribution 
is estimated by parameter free expressions~\cite{Pascalutsa07a}.

Only at low $Q^2$ one can observe some deviation for $G_C^\ast$ below 0.15 GeV$^2$.
This gap between theory and empirical data has been discussed 
in the literature~\cite{Siegert-ND,RSM-Siegert,GlobalFit}.
It was recently shown that the $G_C^\ast$ data have been overestimated.
The new measurements and the recent data analysis evidence a reduction 
in the values for $G_C^\ast$ below 0.15 GeV$^2$~\cite{Blomberg16a}.
The result of a new analysis of JLab/Hall A was 
presented during the workshop by N.~Sparveris~\cite{Sparveris-talk}.


\subsection{Siegert's theorem}

An interesting discussion about the 
$\gamma^\ast N \to \Delta(1232)$ quadrupole form factors 
is raised when we consider 
Siegert's theorem~\cite{MAID1,Buchmann98,MAID2011,Tiator}.
Siegert's theorem states that 
the electric and the Coulomb quadrupole form factors 
are related at the pseudothreshold
when the nucleon and the $\Delta(1232)$ are both at rest,  
by~\cite{Jones73,Siegert1,Siegert2}
\ba
G_E^\ast (Q^2_{pt}) = \frac{M_\Delta-M_N}{2 M_\Delta} G_C^\ast (Q^2_{pt}),
\label{eqST}
\ea
where $Q^2_{pt}= -(M_\Delta -M_N)^2$.
In the following discussion, we  use $\kappa=  \sfrac{M_\Delta-M_N}{2 M_\Delta}$.

When expressed in terms of the scalar ($S_{1/2}$)
and the electric ($E_{1+}$) amplitudes, Eq.~(\ref{eqST})
implies $E_{1+} = \sqrt{2} (M_\Delta -M_N) S_{1/2}/|{\bf q}|$~\cite{Siegert1,Siegert2}.
The previous relation is violated by previous 
MAID parametrizations~\cite{MAID1,MAID2011}.
In Ref.~\cite{Siegert2} it is discussed in detail 
how the helicity amplitudes can be parametrized 
in order to satisfy the condition (\ref{eqST}).
The discussion of  Siegert's theorem can 
also be extended to the $N(1535)$ and $N(1520)$ systems~\cite{Siegert1,Siegert2}.

It was recently shown that the parametrizations 
proposed to estimate the pion cloud effects~\cite{Pascalutsa07a} 
are not fully compatible with Siegert's theorem in 
the form of Eq.~(\ref{eqST})~\cite{Siegert-ND}.
For the purpose of the discussion, 
it is convenient to write 
the pion cloud contributions for the quadrupole form 
factors in the form 
\ba
G_E^\pi = 
 \left(\frac{M}{M_\Delta} \right)^{3/2} 
\frac{M_\Delta^2 -M_N^2}{2 \sqrt{2}} 
\frac{\tilde G_{En}}{ 1 + \alpha},
\hspace{.8cm}
G_C^\pi =
\left(\frac{M_N}{M_\Delta} \right)^{1/2} 
\sqrt{2} M_\Delta M_N \tilde G_{En}, \nonumber \\
\label{eqGEGCst}
\ea
where $\tilde G_{En} = G_{En}/Q^2$, and 
$\alpha$ is a function to be discussed next.
The original form  proposed by Pascalutsa and Vanderhaeghen
has $\alpha \equiv 0$~\cite{Pascalutsa07a}.

To measure the error in  Eq.~(\ref{eqST}), 
we use 
${\cal R}_{pt} \equiv  G_E^\ast (Q^2_{pt}) - \kappa G_C^\ast (Q^2_{pt})$.
An exact description of Eq.~(\ref{eqST}) implies that ${\cal R}_{pt} =0$. 

Starting with the original parametrizations (\ref{eqGEGCst}),
where $\alpha =0$, 
one can conclude using the approximation 
$G_{En} \simeq - \sfrac{1}{6} r_n^2 Q^2$ 
for small values of $Q^2$  (including $Q^2_{pt}$),
that ${\cal R}_{pt} = {\cal O}(1/N_c^2)$.
An error ${\cal R}_{pt} = {\cal O}(1/N_c^2)$ can be significant
in numerical calculations~\cite{Siegert-ND}.

\begin{figure}[t]
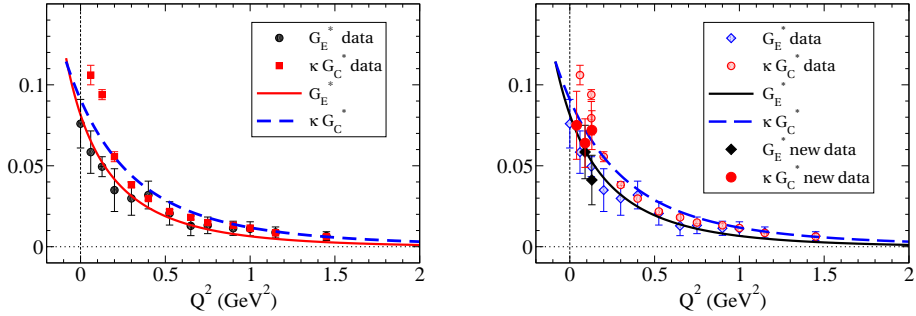

\vspace{.5cm}
\includegraphics[width=5.5cm]{GE-GCmod2c}  \hspace{.8cm}
\includegraphics[width=5.5cm]{GE-GCmod3p-v6} 
\caption{Siegert's theorem.
{\bf At the right:} Results from Ref.~\cite{Siegert-ND}.
{\bf At the left:} Results from Ref.~\cite{RSM-Siegert}. 
Data from Ref.~\cite{MokeevDatabase}.
The new data are from JLab/Hall A (solid circles and solid diamonds)~\cite{Blomberg16a}.}
\label{figGEGC2}       
\end{figure}

It was found that the description of Siegert's theorem 
and the data can be improved when we consider
$\alpha = \frac{Q^2}{M_\Delta^2-M^2_N}$~\cite{Siegert-ND}.
In that case we can write 
${\cal R}_{pt} = (\sfrac{M_\Delta}{M_N})^{\frac{3}{2}} 
\sfrac{M_\Delta -M_N}{2 M_N} \sfrac{r_n^2}{12 \sqrt{2}} Q_{pt}^2$
corresponding to a term ${\cal O}(1/N_c^4)$~\cite{Siegert-ND}, 
a negligible violation of Siegert's theorem.
The results for $G_E^\ast$ and $\kappa G_C^\ast$ 
are presented in the left panel of Fig.~\ref{figGEGC2}. 
The almost convergence of $G_E^\ast$ and $\kappa G_C^\ast$ 
for the lowest value of $Q^2$ is an indication 
of the almost validity of Siegert's theorem.

To the agreement with the data contribute also
the inclusion of the valence quark component, 
as discussed previously relative to the results from Fig.~\ref{figGEGC1}.
The advantage of the estimate of valence quark contribution 
with a covariant quark model 
is that the contributions for the quadrupole form 
factors $G_E^\ast$ and $G_C^\ast$ vanishes at the pseudothreshold, 
as a consequence of the orthogonality between nucleon and 
$\Delta(1232)$ states~\cite{NDeltaD,LatticeD,Siegert-ND}.

The previous result, ${\cal R}_{pt} = {\cal O}(1/N_c^4)$,
was improved in a more recent work~\cite{RSM-Siegert}.
We obtain an exact description of Siegert's theorem 
when we consider $\alpha =  \frac{Q^2}{2 M_\Delta (M_\Delta-M)}$,
which correspond to a correction of the previous representation 
by a term $ {\cal O}(1/N_c^4)$ at the pseudothreshold. 
When combined with the valence quark contribution, 
one obtains again a good agreement with the quadrupole form factor data.
The results for  $G_E^\ast$ and $\kappa G_C^\ast$  are present 
in the right panel of Fig.~\ref{figGEGC2},
in comparison with the data from Refs.~\cite{Blomberg16a,MokeevDatabase}.

The agreement from the  right panel of Fig.~\ref{figGEGC2}
is more impressive because we include the most recent data for $G_C^\ast$
from JLab/Hall A, below 0.15 GeV$^2$ 
(see full circles and full diamonds). 
When we consider the new data we obtain, at last, a consistent description of the 
quadrupole form factor data at low $Q^2$. 

We recall that for this agreement contributes 
the inclusion of a valence quark component 
estimated by a covariant quark model calibrated by lattice QCD data 
(no pion cloud contamination)~\cite{LatticeD} 
and an improved parametrizations of the 
pion cloud contributions to the quadrupole form factors 
with no adjustable parameters.

More recently, the relations (\ref{eqGEGCst}) 
with $\alpha =  \frac{Q^2}{2 M_\Delta (M_\Delta-M_N)}$
have used to study alternative parametrizations
for the neutron electric form factor $G_{En}$~\cite{GlobalFit}.

\subsection{Summary of the $\Delta(1232)3/2^+$ results}

We can now summarize the covariant spectator quark model results for the 
$\gamma^\ast N \to \Delta(1232)$ transition.

We conclude that the simplest model, where 
the nucleon and the $\Delta(1232)$ are described by 
quark-diquark $S$-states, is consistent 
with the results for $G_M^\ast$ from lattice QCD simulations 
and with estimates from the bare core effects 
based on the EBAC/Argonne-Osaka dynamical model.

When we take into account $D$-states in the $\Delta(1232)$
one obtain also a consistent description 
of $G_M^\ast$ and the quadrupole form factors  $G_E^\ast$ and $G_C^\ast$.
$G_M^\ast$ is dominated by valence quark contributions.
The quadrupole form factors are dominated by pion cloud contributions.

The present model for $G_M^\ast$, including 
the parametrization of the pion cloud contribution ($G_M^\pi$)
can be used as input in calculations of other processes.
Examples of that are the extension of the model
to the timelike region 
($\Delta(1232)$ Dalitz decay)~\cite{NDeltaTL,NDeltaTL2}, 
and the octet to decuplet transitions~\cite{OctetDecuplet,OctetDecuplet2}.
The model for the $\Delta(1232)$ structure can also be used 
to calculate $\Delta(1232)$ elastic form factors~\cite{Omega,Deformation,DeltaFF,DeltaDFF}.


\section{Results for the $[70,1^-]$ supermultiplet}
\label{secSQTM}

One can estimate the helicity amplitudes of the 
resonances from the $[70,1^-]$ supermultiplet with the help 
of the single transition quark model (SQTM)~\cite{SQTM,Hey74,Cottingham79,Burkert03}.
According to the $SU(6)\otimes O(3)$ classification the components
of the $[70,1^-]$ supermultiplet are particles 
and angular momentum $J= \frac{1}{2}, \frac{3}{2}$ and  negative parity.
More explicitly, the supermultiplet includes  
the resonances $N(1535)1/2^-$ and   $N(1520)3/2^-$ discussed 
previously in Sect.~\ref{secSRapp}, in addition to 
$N(1650)1/2^-$, $N(1700)3/2^-$,
$\Delta(1620)1/2^-$ and  $\Delta(1700)3/2^-$.

The SQTM assumes that 
in the electromagnetic  interaction the photon 
couples with a single quark, and that the wave functions 
of the baryons are described by the $SU(6) \otimes O(3)$ symmetry group.
In these conditions one can express the quark transverse current
in the form~\cite{Burkert03}
\ba
J^+ = A L^+  + B \sigma^+L_z  +  C \sigma_z L^+ + ...,
\label{eqJp}
\ea
where $L$ is the orbital angular operator, $\sigma$ is the 
Pauli spin operator, and  
$A$, $B$ and $C$ are functions of $Q^2$.
The operator act on the quark spatial wave functions.
The dots represent an extra term relevant for other supermultiplets 
but that is absent for the $[70,1^-]$ supermultiplet.

As a consequence from Eq.~(\ref{eqJp}) 
the transverse helicity amplitudes of the transition $\gamma^\ast N \to N^\ast$,
where $N^\ast$ is a member of the $[70,1^-]$ supermultiplet,
can be represented in terms of the three independent functions 
$A$, $B$ and $C$, and two mixture angles $\theta_S$ and $\theta_D$
determined experimentally~\cite{SQTM,Capstick00,Burkert03}.
The relation between the amplitudes and the coefficients 
can be found in Ref.~\cite{SQTM}.

The coefficients $A$, $B$ and $C$ can be determined 
for the case of proton targets using 
the available data for finite $Q^2$ associated 
with three amplitudes for some resonances $N^\ast$ from the supermultiplet. 
We choose then the 
amplitudes associated with the states $N(1535)1/2^-$
($A_{1/2}^S$) and $N(1520)3/2^-$ ($A_{1/2}^D$ and $A_{3/2}^D$)
calculated within the covariant spectator quark model  
from Refs.~\cite{N1535,N1520,N1520TL}.
One can combine the frameworks from the SQTM 
with the covariant spectator quark model because both frameworks are based 
on the $SU(6)$ wave functions and on the impulse approximation.
However, since both frameworks are restricted 
to the valence quark degrees of freedom we cannot 
expect the estimates to hold for small $Q^2$.

In Refs~\cite{SQTM,N1535,N1520,N1520TL} the valence quark contributions 
to the transition form factors and helicity amplitudes 
for $N(1535)1/2^-$ and $N(1520)3/2^-$ 
are calculated in terms of the radial wave functions of the resonance.
Adjusting just one parameter of the radial wave functions 
per resonance, associated with the long range behavior, 
one can reproduce the helicity amplitude data for $Q^2 > 2$ GeV$^2$,
with a minor exception.
The amplitude $A_{3/2}$ for the state $N(1520)3/2^-$ cannot 
be described in the context of the covariant spectator quark model, which predicts $A_{3/2}^D=0$,
as mentioned in Sect.~\ref{secSRapp}.
As discussed, the result $A_{3/2}^D=0$ can be interpreted as a manifestation
that the amplitude $A_{3/2}^D$ is dominated by 
meson cloud effects~\cite{SQTM,N1520,N1520TL}.

\begin{figure}
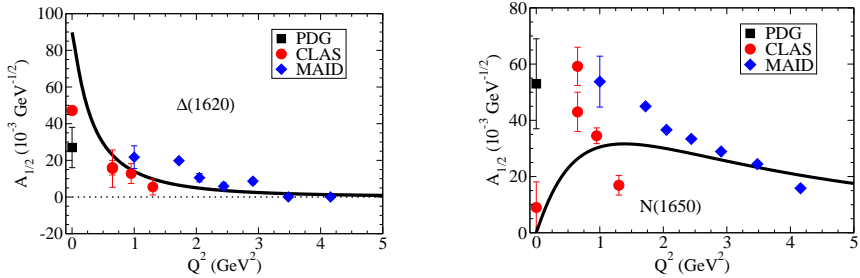

\vspace{.5cm}
\includegraphics[width=5cm]{D1620Z1}  \hspace{1.cm}
\includegraphics[width=5cm]{N1650aZ1} 
\caption{Results from the parametrization from SQTM.
{\bf At left:} $\Delta(1620)1/2^-$.
{\bf At right:} $N(1650)1/2^-$. 
Data from Refs.~\cite{MAID1,Aznauryan09,Mokeev12,PDG,Aznauryan05,Dugger09}.}
\label{figSQTM}       
\end{figure}

The interpretation of $A_{3/2}^D$ as the representation 
of the meson cloud contributions can be used 
to identify the meson cloud contributions 
to the remaining amplitudes of the supermultiplet,
tracking the contribution of $A_{3/2}^D$ on those amplitudes.
By subtraction one can also identify 
the valence quark contributions~\cite{SQTM}.

In Fig.~\ref{figSQTM}, we present the results for 
the amplitudes associated with the states $N(1650)1/2^-$ 
and $\Delta(1620)1/2^-$, 
in comparison with the data from CLAS, MAID and PDG.
Predictions for the resonances $N(1700)3/2^-$ and  $\Delta(1700)3/2^-$ 
are presented in Ref.~\cite{SQTM}.
We do not discuss those results here because 
the data for $Q^2 > 1$ GeV$^2$ are scarce or nonexistent~\cite{SQTM}.
In Fig.~\ref{figSQTM}, one can observe a good general agreement 
between the estimate based on the SQTM and the data.
There is, however, an important difference between the two results.
The $N(1650)1/2^-$ amplitude is independent of the parametrization $A_{3/2}^D$,
meaning that the result is the same when we drop  the contribution of 
that amplitude ($A_{3/2}^D \to 0$).
As for the $\Delta(1620)1/2^-$ amplitude, the contributions 
associated with the amplitudes 
$A_{1/2}^S$ and $A_{1/2}^D$ cancel almost exactly.
One concludes, then that $A_{1/2} \propto A_{3/2}^D$,
meaning that the SQTM cannot predict   $A_{1/2}$ 
for $\Delta(1620)1/2^-$, unless the meson cloud effects are included explicitly.
For a more detailed, discussion check Ref.~\cite{SQTM}.
Since the data available for the $\Delta(1620)1/2^-$ for 
large $Q^2$ are restricted to the MAID data, which 
have abnormally small errorbars, it will be very 
interesting to see if the future data from the JLab-12 GeV upgrade 
confirms or deny the present trend of the calculations.
The same observation holds for $N(1650)1/2^-$ amplitude 
and for the amplitudes associated with 
the resonances $N(1700)3/2^-$ and $\Delta(1700)3/2^-$,
discussed in Ref.\cite{SQTM}.

The results of SQTM framework based on the  covariant spectator 
quark model results from Refs.~\cite{N1535,N1520,N1520TL} 
with adjustable radial wave function $\psi_R$ 
can in a near future be improved based on the covariant spectator quark model 
results in the semirelativistic limit (Sect.~\ref{secSRapp}).
In that case, we can obtain parametrizations 
to the $[70,1^-]$ supermultiplet which depend 
only on the nucleon parametrization for the radial wave function $\psi_N$.

\section{Summary and conclusions}
\label{secConclusions}

We present here covariant estimates for the  
transition form factors and the helicity amplitudes 
for several nucleon excitations $N^\ast$.
We discussed, in particular the results for 
the resonances $\Delta(1232)3/2^+$, $N(1440)1/2^+$ 
$N(1535)1/2^-$ and $N(1520)3/2^-$ as 
well as results for the resonances of 
$[70,1^-]$ supermultiplet,  $\Delta(1620)1/2^-$ and 
$N(1650)1/2^-$, based on a connection 
with the single quark transition model.

In general, we observed a good agreement between model predictions 
and empirical data at large $Q^2$.
This result can be interpreted as an indication 
that the effects of the valence quark degrees of freedom 
in the model are under control. 

For the $\Delta(1232)3/2^+$ there is today a convergence 
of results from different methods, 
including lattice QCD simulations, quark models, 
Dyson-Schwinger equations and also from dynamical 
coupled-channel models.

As for the other resonances, 
the estimates based on different methods are still under debate.
At the moment, the best that we can do, 
for sure, is to make predictions for large $Q^2$.

Based on the results obtained within the framework of the covariant 
spectator quark model, 
we can conclude that the meson cloud effects may be relevant at moderated $Q^2$ 
(region around $Q^2 \approx 2 $ GeV$^2$) 
for some resonances $N^\ast$.
Here we discussed the Pauli form factor for the resonance $N(1535)1/2^-$,
and the amplitude $A_{3/2}$ for the  resonance $N(1520)3/2^-$.

In the near future we can expect developments related 
to the following topics:
\begin{itemize}
\item
New data at large $Q^2$ and more precise data at any range 
can help the interpretation of the empirical results.
In this context the coming results from 
the JLab-12 GeV-upgrade may be particularly relevant.
\item
Lattice QCD simulations below the $N^\ast$ threshold 
will help to refine the interpretation of our theoretical models
based on valence quark and meson cloud degrees of freedom.
\item
New estimates of the bare core contributions to 
the transition form factors based on dynamical couple-channel models
and the comparison with estimates from quark models 
can help to understand the data for several resonances $N^\ast$.
This comparison was already very useful in the past 
for the $\Delta(1232)3/2^+$, as discussed in the present work.
\end{itemize}

\begin{acknowledgements}
The author thanks Adolf Buchmann, Ralf Gothe, Viktor Mokeev, Beatrice Ramstein,
Elena Santopinto and Nikolaos Sparveris for helpful discussions
and Witold Przygoda and the HADES collaboration the results from Ref.~\cite{HADES17} 
(right panel of Fig.~\ref{figGMTL}).
This work was supported by the Funda\c{c}\~ao de Amparo \`a 
Pesquisa do Estado de S\~ao Paulo (FAPESP):
project no.~2017/02684-5, grant no.~2017/17020-BCO-JP. 
\end{acknowledgements}





\end{document}